\documentclass[a4paper,fleqn]{cas-sc}
\usepackage{amssymb, amsmath}
\usepackage[numbers]{natbib}
\usepackage{graphicx} % 
\usepackage{subfig}

\def\tsc#1{\csdef{#1}{\textsc{\lowercase{#1}}\xspace}}
\tsc{WGM}
\tsc{QE}
\tsc{EP}
\tsc{PMS}
\tsc{BEC}
\tsc{DE}
\title [mode = title]{Correlation between microstructural deformation mechanisms and acoustic parameters on a cold-rolled Cu30Zn brass}

\begin{document}
\let\WriteBookmarks\relax
\def\floatpagepagefraction{1}
\def\textpagefraction{.001}

\shorttitle{Acoustic parameters on a cold-rolled Cu30Zn brass}
\shortauthors{M. Sosa, et~al.}

\author[1]{Maria Sosa}
\cormark[1]
\affiliation[1]{organization={Departamento de Ingeniería Metalúrgica, Universidad de Santiago de Chile },
    addressline={Av. Ecuador 3735}, 
    city={Santiago de Chile},
    % citysep={}, % Uncomment if no comma needed between city and postcode
    postcode={9170124}, 
    % state={},
    country={Chile}}
 
% Second author
\author[1]{Linton Carvajal}

% Third author
\author[2]{Vicente Salinas}[orcid=0000-0002-5759-9371]
\cormark[2]

% Fourth author
\author%
[4] {Fernando Lund}

% Fifth author
\author%
[3] {Claudio Aguilar}

% Sixth author
\author%
[1] {Felipe Castro}[orcid=0000-0001-8126-5593]

\affiliation[2]{organization={Grupo de Investigación Aplicada en Robótica e Industria 4.0, Facultad de Ingeniería, Universidad Autónoma de Chile},
addressline={Av. Pedro de Valdivia 641}, 
city={Santiago},
 % citysep={}, % Uncomment if no comma needed between city and postcode 
country={Chile}}

\affiliation[3]{organization={Departamento de Ingeniería Metalúrgica y Materiales, Universidad Técnica Federico Santa Maria},
    addressline={Av. España 1680}, 
    city={Valparaiso}, 
    country={Chile}}
    
\affiliation[4]{organization={Departamento de Física, Facultad de Ciencias Físicas y Matemáticas, Universidad de Chile},
    addressline={Avenida Blanco Encalada 2008}, 
    city={Santiago de Chile},
    country={Chile}}

% Corresponding author text
\cortext[cor1]
 {maria.sosa.o@usach.cl}

 \cortext[cor2]
 {vicente.salinas@uautonoma.cl}

\begin{abstract}
    The relationship between acoustic parameters and the microstructure of a Cu30Zn brass plate subjected to plastic deformation was evaluated. The plate, previously annealed at 550 °C for 30 minutes, was cold rolled to reductions in the 10-70\% range. Using the pulse-echo method, linear ultrasonic measurements were performed on each of the nine specimens, corresponding to the nine different reductions, recording the wave times of flight of longitudinal wave along the thickness axis. Subsequently, acoustic measurements were performed to determine the nonlinear parameter ($\beta$) through the second harmonic generation. X-ray diffraction analysis revealed a steady increase and subsequent saturation of deformation twins at 40\% thickness reduction. At higher deformations, the microstructure revealed the generation and proliferation of shear bands, which coincided with a decrease in the twinning structure and an increase in dislocation density rate. Longitudinal wave velocity exhibited a 0.9\% decrease at 20\% deformation, followed by a continuous increase of 2\% beyond this point. These results can be rationalized as a competition between a proliferation of dislocations, which tends to decrease the linear sound velocity, and a decrease in average grain size, which tends to increase it. These variations are in agreement with the values obtained with XRD, Vickers hardness and metallography measurements. The nonlinear parameter $\beta$ shows a significant maximum, at the factor of 8 level, at 40\% deformation. This maximum correlates well with a similar maximum, at a factor of ten level and also at 40\% deformation, of the twinning fault probability.

\end{abstract}

% Keywords

\maketitle

\section{Introduction}
Alpha brass (or yellow brass) is a copper-brass alloy with less than 35\% Zinc, in which the latter is dissolved to form a solid solution of uniform composition. The best-known alpha brass is the Cu30Zn or cartridge brass, which has the optimum combination of properties such as a yield strength of 76 to 448 MPa, a tensile strength of 303 to 896 MPa and high ductility with elongations up to 66\%, making it suitable for cold work \cite{ asm1990properties}.

The principal mechanisms of plastic deformation in face-centered cubic (FCC) metals and alloys consist of slip via dislocation motion and deformation twinning. The Cu30Zn has a FCC structure and stacking fault energy of 14 $mJ/m^{2}$ \cite{Gallagher1970TheIO}, which gives it the ability to form deformation twins and results in high strain hardening. Twinning breaks down a material’s microstructure into progressively smaller domains, a similar effect to grain refining, known as the dynamic Hall-Petch-effect which results in strengthening \cite{meyers2008mechanical}.  

The critical stress needed for twinning is higher than dislocations slip, and the nucleation of deformation twins also requires prior dislocation activity \cite{el2000deformation, el1999influence, fargette1976plastic, hirsch1985orientation, hutchinson1979development, madhavan2015role, duggan1978deformation}. Studies \cite{el2000deformation, el1999influence} have shown that the onset of twinning on Cu30Zn during uniaxial tensile and compression tests is accompanied by a change in the slope in a $d\sigma/d\varepsilon$-type plot \cite{Kocks2003PhysicsAP}, an inflexion point associated with critical twinning stress. During cold rolling reduction, it has been shown  \cite{fargette1976plastic, hirsch1985orientation, hutchinson1979development, madhavan2015role, duggan1978deformation} that twins are a prominent feature of the microstructure up to 40-50\% reduction while at about 50-60\% reduction, shear bands appear. These bands correspond to narrow zones of intensified shearing strain that cut across grains.

Ever since the classic work of Fargette and Whitham \cite{fargette1976plastic} and Duggan et al. \cite{duggan1978deformation}, it has been known that the microstructure of alpha brass, steels, and metallic materials, in general, undergoes complexity during large deformations \cite{Raphanel}. This complexity has been the subject of extensive research. Recent examples of experimental studies include the examination of TWIP steels using Optical, Scanning Electron Microscopy (SEM), Transmission Electron Microscopy (TEM), and Electron Backscattering Diffraction (EBSD) at 75\% deformation \cite{gupta}. In another study, a high entropy alloy was deformed up to a shear strain of 11 and analyzed using EBSD, TEM, X-ray diffraction (XRD), and plasticity simulations \cite{KUMARAN2023118814}. Similarly, a Ni-Ti alloy at 40\% strain was investigated, employing TEM, SEM, and XRD \cite{CHEN2022142136}. Commercial stainless steel underwent an 87\% reduction in thickness, and this transformation was analyzed using electron microscopies and XRD \cite{CHEN2022143224}. Neutron diffraction and EBSD were utilized to study $TiB_{2}/Al$  composites with deformations of up to 95\% thickness reduction \cite{DAN2018293}. 

The various microscopies mentioned in the last paragraph, as well as neutron and X-ray diffraction, are probes that have been developed to a very high degree of sophistication. Nevertheless, the fact remains that they are sensitive only to surface phenomena, or are quite intrusive, in the sense that they do not probe the properties of a material itself, but of a material that has been specially prepared to fit the requirements of a microscope, or a neutron reactor, or a diffractometer. By contrast, ultrasound penetrates well into the bulk of a material and, due to of the very low energies involved, is not intrusive. Because of these features, it lends itself to the probing of pieces in service. 

Linear or conventional ultrasonic techniques have been used as a nondestructive technique for detecting defects such as cracks, pores, inclusions, and thickness loss \cite{burhan2019guideline, shao2005review, merazi2010automatic, d2008automatic}. These techniques, mainly associated with time of flight, echoes detection and attenuation measurement, are highly sensitive to the millimeter to micron size ranges defects but less sensitive to features of a lower order of magnitude  \cite{jhang2009nonlinear, kniazevnumerical, Buck1990, cantrell2004fundamentals}. Although wave velocity measurement can be used to detect changes in microstructure associated to heat treatments or deformation \cite{carvajal2021ultrasonic}, for such microstructural features as lattice distortion, crystallographic orientation, grain size variation, and microstructural defect density, wave distortion monitoring through nonlinear ultrasonic techniques provides a competitive material characterization technique \cite{granato1956theory, granato1956application, hikata1965dislocation, hikata1966generation, cantrell1997effect, cantrell1998nonlinear, cantrell2006quantitative, nazarov1997nonlinear, 166766, mini2015experimental, li2019characterization, viswanath2011nondestructive, espinoza2018linear}.   

A nonlinear acoustic experimental method that has been shown capable of detecting and monitoring microstructural changes is the second harmonic generation (SHG) in metals \cite{matlack2015review}. In this method, a second harmonic wave is generated from the propagation and interaction of a monochromatic elastic wave with the medium. As the wave passes through the medium, it is distorted by microstructural features, which results in harmonics generation. This nonlinear response is quantified through the nonlinear parameter $\beta$, which is formally defined in terms of linear combinations of second and third-order elastic constants. However, since this parameter is part of the solution to the nonlinear wave equation, it is reduced to a proportionality factor between the amplitudes of the first and second harmonics, magnitudes that can be experimentally measured \cite{matlack2015review}. 

Phenomena such as slip dislocation, precipitation hardening, martensitic transformations, fatigue microcrack formation, and grain size are those whose nonlinear effect has been evaluated through harmonic generation. This harmonic generation arises, for example, from the movement and multiplication of dislocations, due to the deviation of elasticity as plasticity arises  \cite{granato1956theory, granato1956application, hikata1965dislocation, hikata1966generation}. During ageing, the nonlinear effect is attributed to the matrix-precipitate coherence that develops due to the effect of heat treatment time and temperature \cite{cantrell1997effect, cantrell1998nonlinear}; in the study of fatigue microcrack formation, the nonlinearity is related to the formation of complex dislocation structures generating favorable places for microcracks nucleation \cite{cantrell2006quantitative, nazarov1997nonlinear}. The increase in nonlinearity due to the decrease in grain size can be attributed to the increase in grain-boundary area and its effect on waveform distortion \cite{mini2015experimental}. 

In a recent work \cite{salinas2022situ}, dislocation density was monitored continuously in a standard tension test for a 304L steel, using ultrasound. Three consecutive loading-unloading cycles were performed, measuring the shear wave velocity, applied force, and sample strain and modelling the stress, strain, and sample thickness through finite elements. The resulting data for dislocation density and acoustic wave velocities as a function of stress in the first and third cycles agree with the predictions of Maurel et al. theory \cite{PhysRevB.72.174110, PhysRevB.72.174111}  which have been independently verified on aluminum results \cite{salinas2017situ, espinoza2018linear}, in which shear wave velocity showed a decrease at the onset of plasticity due to dislocation proliferation. The plastic behavior in the second cycle is markedly different, with an increase, rather than a decrease, in shear wave velocities. According to X-ray diffraction (XRD) measurements and a Croussard–Jaoul (CJ) analysis of the stress-strain curve, the change in wave velocity was attributed to the presence of an increasing number of twins in this intermediate cycle and it was shown, using the model of Hirsekorn \cite{10.1121/1.388233, 10.1121/1.389206}, that this wave velocity increase is consistent with a decrease in grain size due to a proliferation of twins.

Microstructural changes that take place in Cu30Zn brass during cold rolling are expected to cause disturbances in the propagation of ultrasonic waves. However, it is found that the wave interaction with twins has been little explored. The question arises as to how linear and nonlinear parameters respond to this deformation mechanism and how sensitive they are to the microstructural variations generated. Therefore, in this study, we present wave propagation velocities and the nonlinear parameter for a Cu30Zn brass cold rolled up to 70\% thickness reduction, relating the results to metallographic observations and results obtained in hardness and x-ray diffraction tests.

\section{Materials and Methods} \label{2}
A Cu30Zn Brass (UNS C26000) plate was homogenized at 550 °C for thirty minutes in a resistance furnace. Subsequently, cold rolling was carried out to obtain nine samples for thickness reductions of up to 70\%, one for each applied deformation. The chemical composition and thickness are shown in Tables 1 and 2, respectively. 

\begin{table}[width=.9\linewidth,cols=4,pos=h]
\begin{minipage}{\textwidth}
\renewcommand{\thefootnote}{\thempfootnote}
\caption{CuZn30 Brass chemical composition, wt\%. {Measured by optical emission spectrometry-SPECTROMAXx arc/spark metal analyzer }}
%\footnote{Measured by optical emission spectrometry-SPECTROMAXx arc/spark metal analyzer }}
\label{tbl1}
\begin{tabular*}
{\tblwidth}{@{} LLLLL@{} }
\toprule
\%Cu & \%Zn &\%Fe & \%P & \%Pb \\
\midrule
69.1 & 30.76 & 0.026 & 0.029 & <0.001 \\
\toprule
\%Mn & \%Ni &\%Si & \%Cr & \%Sn \\
\midrule
<0.0005 & 0.0021 & <0.0005 & 0.0018 & <0.0005 \\
\bottomrule
\end{tabular*}
\end{minipage}
\end{table}

\begin{table}[width=.9\linewidth,cols=4,pos=h]
\begin{minipage}{\textwidth}
\renewcommand{\thefootnote}{\thempfootnote}
\caption{Thickness and deformation of cold-rolled samples.} 
\label{tbl2}
\begin{tabular*}
{\tblwidth}{@{} LLL@{} }
\toprule
Sample & Deformation [\%] & Thickness [mm]\footnote{Measured with Mitutoyo-Micrometer EXT. 0-25 mm (0.01mm (103-137).}\\
\midrule
P00 & 0.0 & 12.13 ± 0.01 \\
P10 & 10.85 & 10.81 ± 0.01 \\
P15 & 14.76 & 10.34 ± 0.02 \\
P20 & 19.48 & 9.77 ± 0.02 \\
P30 & 28.85 & 8.63 ± 0.01 \\
P40 & 39.16 & 7.38 ± 0.01 \\
P50 & 49.43 & 6.13 ± 0.01 \\
P60 & 60.74 & 4.76 ± 0.02 \\
P70 & 69.82 & 3.66 ± 0.03 \\
\bottomrule
\end{tabular*}
\end{minipage}
\end{table}

\subsection{Optical microscopy and hardness test}\label{2.1}

Metallographic analyses and hardness measurements were performed on the RD-ND plane (see Fig. 1). Samples were prepared using standard grinding procedures, followed by polishing with 1, 0.3, 0.05 $\mu m$ alumina and chemical polishing intervals with a copper polishing solution (50 ml acetic acid, 40 ml nitric acid, 10 ml orthophosphoric acid and 1ml hydrochloric acid). Finally, samples were polished with colloidal silica for 30 minutes and over-etched with a 5gr $FeCl_{3}$, 10 ml $HCl$ and 100 ml $H_{2}O$ solution.

Vickers hardness was conducted by a Zwick/Roell Vickers microhardness tester. The hardness measurements were carried out under 300 gf load, making ten indentations per sample. 

By some estimations, Sidor et al. \cite{met11101571} have proven that the hardness test is a useful tool to estimate dislocation density. By approximating the flow stress $\sigma_{y}$ from the Vickers hardness $H_{V}$ ($\sigma_{y}=H_{V}/3.06$), the stored energy ($E_{D}$) caused by the presence of dislocations can be estimated from the measured $H_{V}$  values via the following relationship \cite{DAN2018297, SALEH2018620}:

\begin{equation}
E_{D}=\frac{H_{V}^{2}}{G[3.06M\alpha]^{2}}
\end{equation}

where M is the Taylor factor, $\alpha$ is the geometric constant and G is the shear modulus. The stored energy and the dislocation density $\rho$ are related to each other as \cite{anthonyrecrystallization}:

\begin{equation}
E_{D}=\alpha \rho G {b^{2}}
\end{equation}

By combining Equations (1) and (2), one can estimate the dislocation density by using Equation (3):

\begin{equation}
\rho_{H_{V}}=\frac{1}{\alpha^{3}}[\frac{H_{V}}{3.06GM\alpha}]^{2}
\end{equation}

Although it is an approximative method, Sidor et al. \cite{met11101571} have obtained good results on aluminum samples when comparing the technique with the analysis of dislocations through XRD.

According to the texture analysis of Cu30Zn brass by Duggan et al. \cite{duggan1978deformation}, the copper-texture component {211} <111> remains almost constant initially but then is largely depleted between 40 – 60\% thickness reduction, range in which brass {110}<112> and Goss {110}<001> texture are distinguished until a peak at 95\% reduction is reached. Given the analysis mentioned above, the Taylor factors M used were those reported for Cooper and Brass-Goss texture 3.64 and 2.45, respectively \cite{HONG2003133}, the Burgers vector b = 0.26 nm \cite{dieter1976mechanical}, shear modulus G = 40 GPa \cite{asm1990properties} and  $\alpha =0.5$ was employed for all deformations evaluated.

\subsection{XRD Characterization}\label{2.2}

X-ray powder patterns of the samples were collected on a multipurpose powder diffractometer STOE STADI MP equipped with a DECTRIS MYTHEN 1K detector and using pure Cu K$\alpha$ radiation ($\lambda=1.54056 \textup{~\AA}$, curved Germanium (111) monochromator of the Johann-type). XRD patterns were obtained between 40° and 100° in 2$\theta$, with a step size of 0.10° and a holding time of 6 s per step. The microstructural parameters were obtained by the Rietveld method, using the Materials Analysis Using Diffraction (MAUD) software  \cite{lutterotti1990simultaneous, lutterotti1999maud}, and $LaB_{6} (a=4.1565915  \textup{~\AA})$ as external standard for determining instrumental broadening. To carry out the microstructure analysis, the profile fitting was performed by considering the Delft line broadening model \cite{de1982use, delhez1993crystal} and anisotropic size-strain model implemented using the Popa’s rule \cite{popa1998hkl}, harmonic texture \cite{de1982use, delhez1993crystal} and planar defects (stacking fault and twins) by Warren’s model \cite{warren2012x}. Cell parameter, microstrain, crystallite size and twin fault probability (TFP) were extracted from the Rietveld refinements and the deviation was calculated by varying the initial microstrain and crystallite size parameters for three refinements of the same condition. The data extracted from the Rietveld refinements were also used to determine the dislocation density, $\rho_{XRD}$, using

\begin{equation}
\Lambda=\frac{\pi \langle \varepsilon^{2}\rangle^{1/2}}{b^{2}}
\end{equation}

where $\langle\varepsilon^{2}\rangle^{1/2}$ is the microstrain, $D$ is the crystallite size (both values are directly obtained from MAUD post-refining) and $b$ is the Burgers vector \cite{bhaskar2014mechanical}, which has been estimated to be 0.26 nm for FCC brass \cite{dieter1976mechanical}. 

\subsection{Acoustic Measurements}\label{2.3}

\subsubsection{Linear acoustic measurements}\label{2.3.1}

5 MHz central frequency longitudinal wave pulses generated by a Panametrics-5077PR pulse generator were induced normal to the RD-TD plane (see Fig. 1) on each deformed specimen using the pulse-echo technique and a contact piezoelectric transducer (V406 -Olympus longitudinal wave with element diameter 13mm). The signals of the wave traveling along the thickness (the ND direction) of the samples were acquired by a Handyscope HS6 oscilloscope and analyzed by TiePie's Multi-Channel software. To determine longitudinal wave velocity, the time of flight were measured as the peak-to-peak time between the first two consecutive echoes. To obtain a standard deviation, measurements were taken at five different points on the RD-TD plane. As specified in Figure 1, $V_{L,N}$  is the longitudinal wave velocity, which is polarized along the propagation direction ND. 

\begin{figure}[ht!]
	\centering
		\includegraphics[scale=.75]{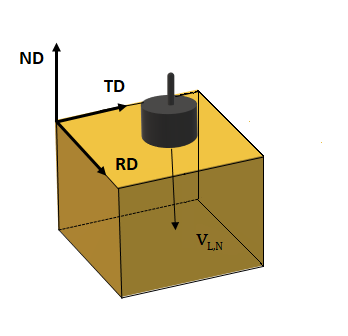}
	\caption{Longitudinal wave propagating along normal (ND) axis. (RD)and (TD) are rolling and transverse axes respectively.}
	\label{FIG:1}
\end{figure}

\subsubsection{Nonlinear acoustic measurements}\label{2.3.2}

Just as for the linear acoustic measurements, the plane of incidence for the fundamental and second harmonic voltage analyses of the deformed brass samples was the RD-TD plane, and the measurements were carried out following the experimental setup shown in Figure 2. Using an Agilent 33250A function generator and an NF - HSA4011 amplifier, a continuous longitudinal wave of frequency 3 MHz was transmitted into the material by a 12.7 mm element diameter Olympus—V109 transducer (resonant at 5 MHz) placed on one side of the sample. The wave was received by an identical transducer and the signal was sent to a Handyscope HS6 oscilloscope connected to a personal computer. Then, through a Fourier analysis, the signal was transferred to the frequency domain, and the voltage of the fundamental harmonic ($V_{\omega}$) and the second harmonic ($V_{2\omega}$) were recorded. For data collection, an amplitude sweep was performed on the emitted wave, between 2.5 and 3 [V], and amplified by a factor of 10. The values of $V_{\omega}$ and $V_{2\omega}$ were obtained for each excitation amplitude. This procedure was executed for nine different points of each sample to test reproducibility and get scatter values. 
 
The nonlinear parameter can be quantified by \cite{matlack2015review}: 
\begin{equation}
\centering
\beta=\frac{8 A_{2\omega}}{x\kappa A_{\omega}^{2}}
\end{equation}

where $x$ is the elastic wave propagation distance[m], $\kappa$ is the wave number [1/m], and $A_{\omega}$ and $A_{2\omega}$ are the absolute physical displacements of the fundamental and second harmonic waves [m]. Since reaching high accuracy on the calibration of the transducers is difficult and acquiring absolute displacements is quite complex, instead of calculating $\beta$ in dimensionless form, $\beta^{*}$  is determined from:

\begin{equation}
\centering
\beta^{*}=\frac{8 V_{2\omega}}{x\kappa V_{\omega}^{2}} [m/V]
\end{equation}

Where $V_{\omega}$ and $V_{2\omega}$ are the amplitudes of the fundamental and second harmonics, measured in voltage units.

\begin{figure}[ht!]
	\centering
		\includegraphics[scale=.75]{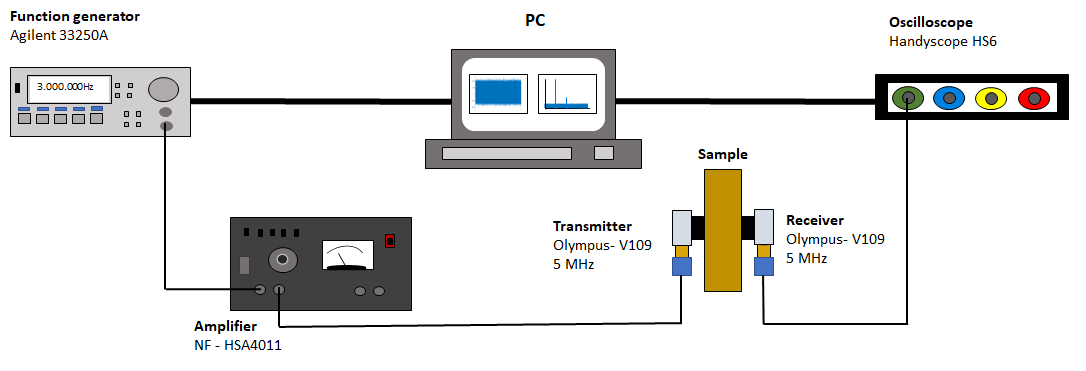}
	\caption{Schematic illustration of the experimental setup used for second harmony generation.}
	\label{FIG:2}
\end{figure}

\section{Results} \label{3}

\subsection{Optical microscopy and hardness results}\label{3.1}

Figure 3 shows the evolution of the microstructure with the deformation. The observation plane is the RD-ND plane, with the rolling direction being the horizontal axis of the micrographs. Figure 3a shows the microstructure of the alpha brass in the softened condition (fully annealed), with a grain size of $\approx 45 \mu m$, calculated by the planimetric procedure \cite{american2004astm}. As the deformation increases, grains become elongated along the rolling direction, and the number of strain marks increases, which are already visible at 10\% thickness reduction (Fig. 3b). These strain marks (indicated by dashed arrows) are visualized as groups of parallel lines within the grain, which intersect with other groups of lines.  The micrographs also show the appearance of shear bands (indicated by solid arrows) after 40\% deformation. These bands are identified as those following an angle between 35°and 45° to the rolling direction, associated with the main shear stresses \cite{duggan1978deformation}.

\begin{figure}[ht!]
\centering
\includegraphics[width=.28\textwidth]{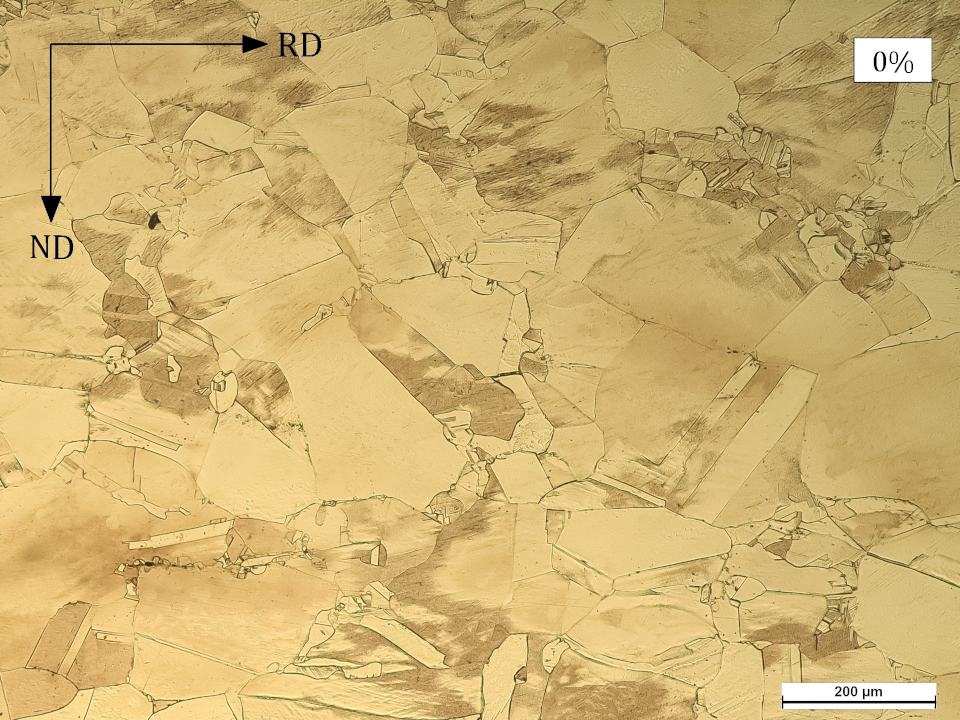}
\label{FIG:3.a}
\includegraphics[width=.28\textwidth]{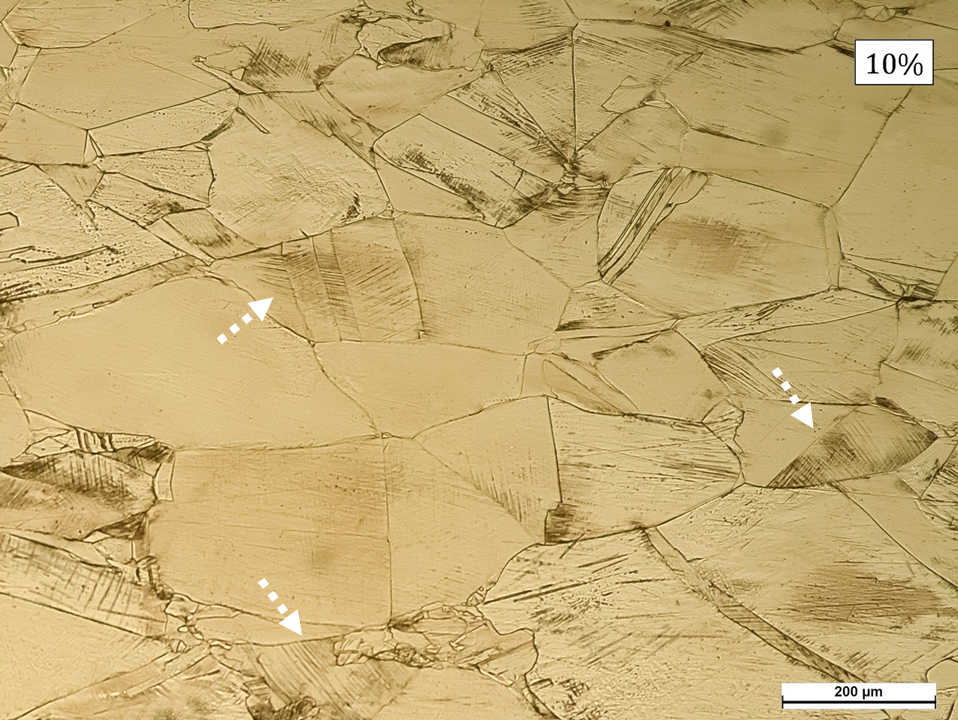}
\label{FIG:3.b}
\includegraphics[width=.28\textwidth]{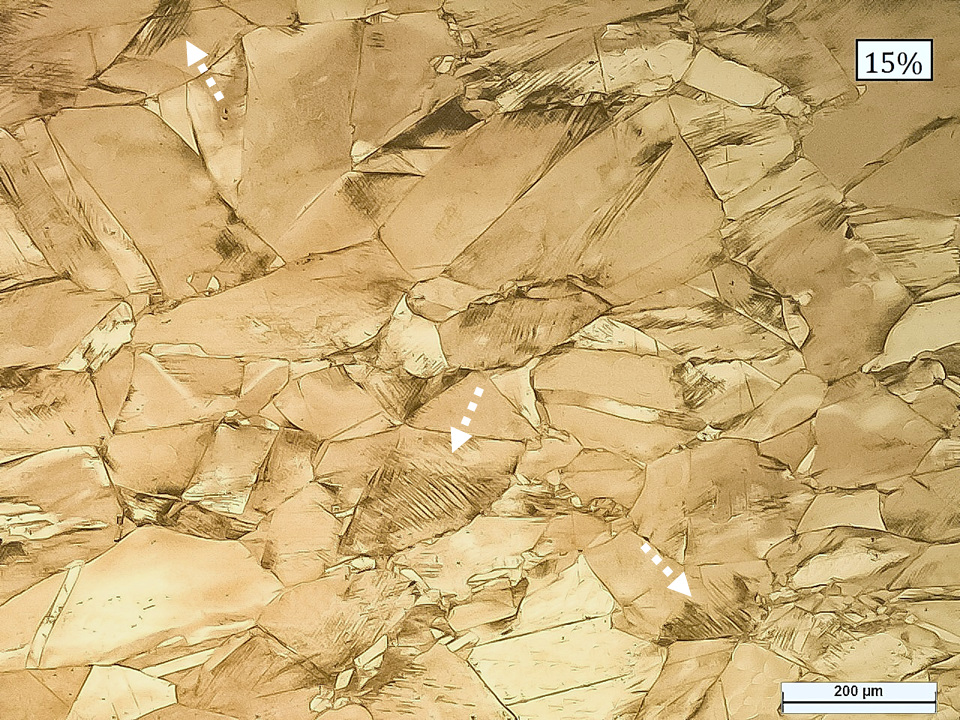}
\label{FIG:3.c}
%\caption{...}
%\end{figure}

%\begin{figure}[ht!]
%\centering
\includegraphics[width=.28\textwidth]{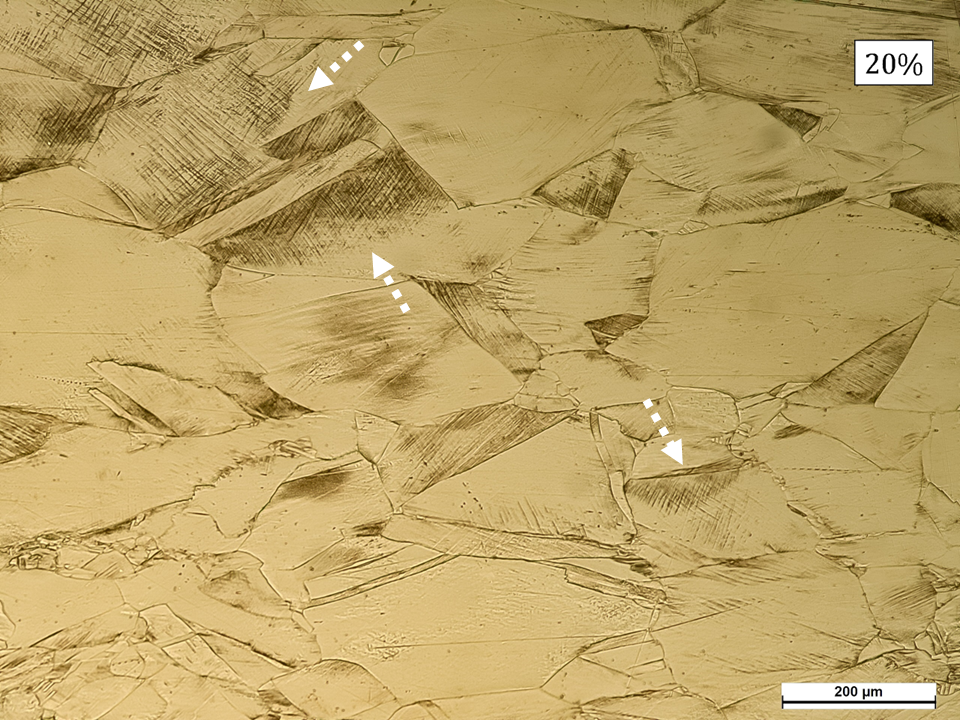}
\label{FIG:3.d}
\includegraphics[width=.28\textwidth]{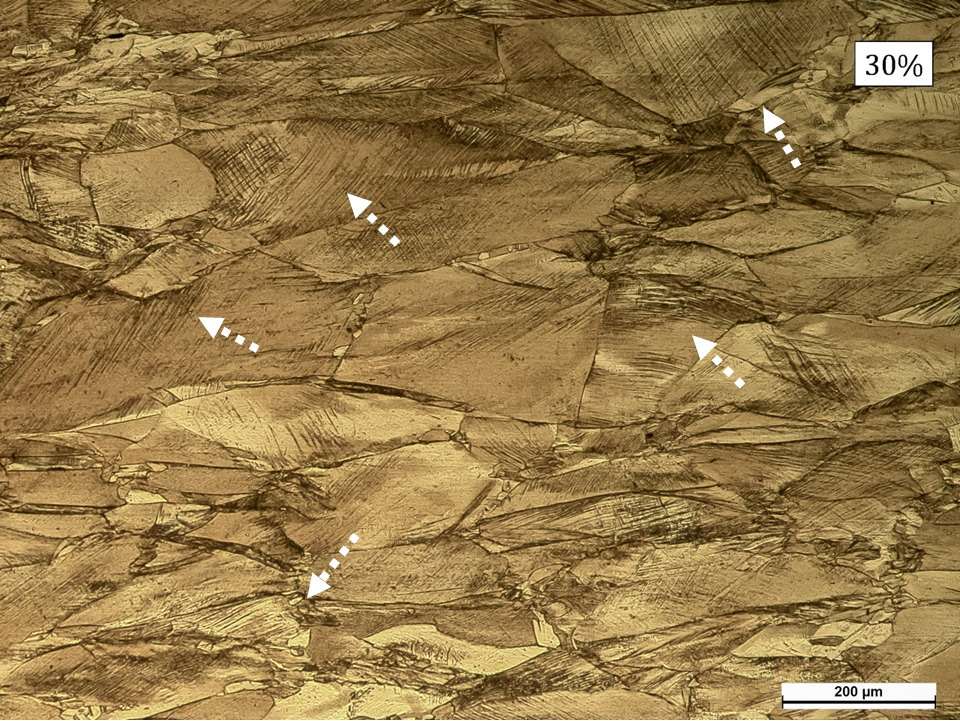}
\label{FIG:3.e}
\includegraphics[width=.28\textwidth]{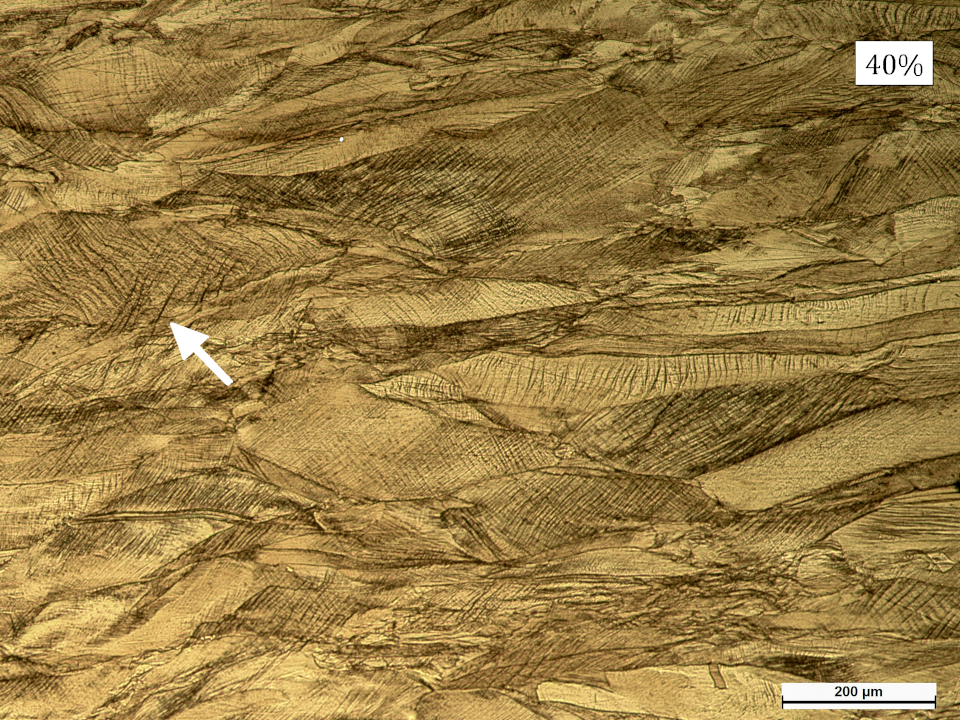}
\label{FIG:3.f}
%\caption{...}
%\end{figure}

%\begin{figure}[ht!]
%\centering
\includegraphics[width=.28\textwidth]{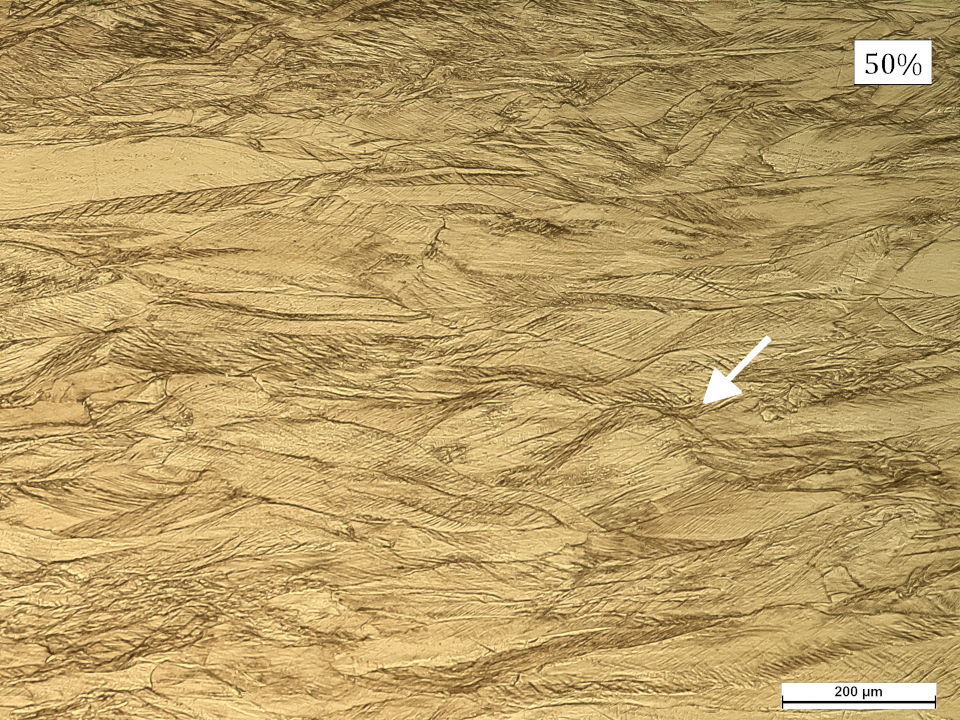}
\label{FIG:3.g}
\includegraphics[width=.28\textwidth]{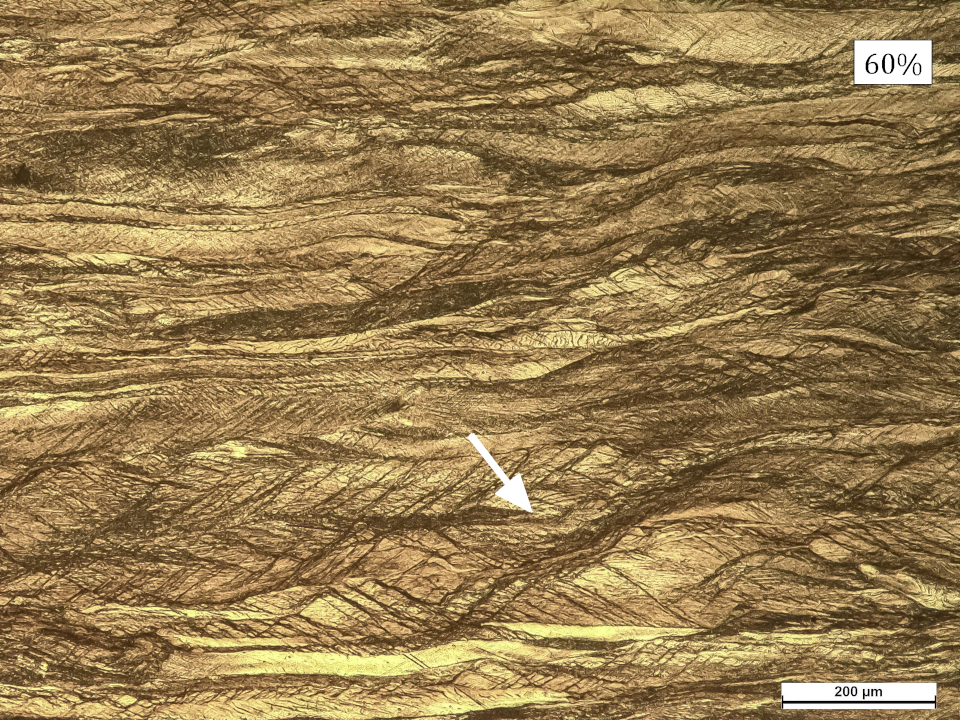}
\label{FIG:3.h}
\includegraphics[width=.28\textwidth]{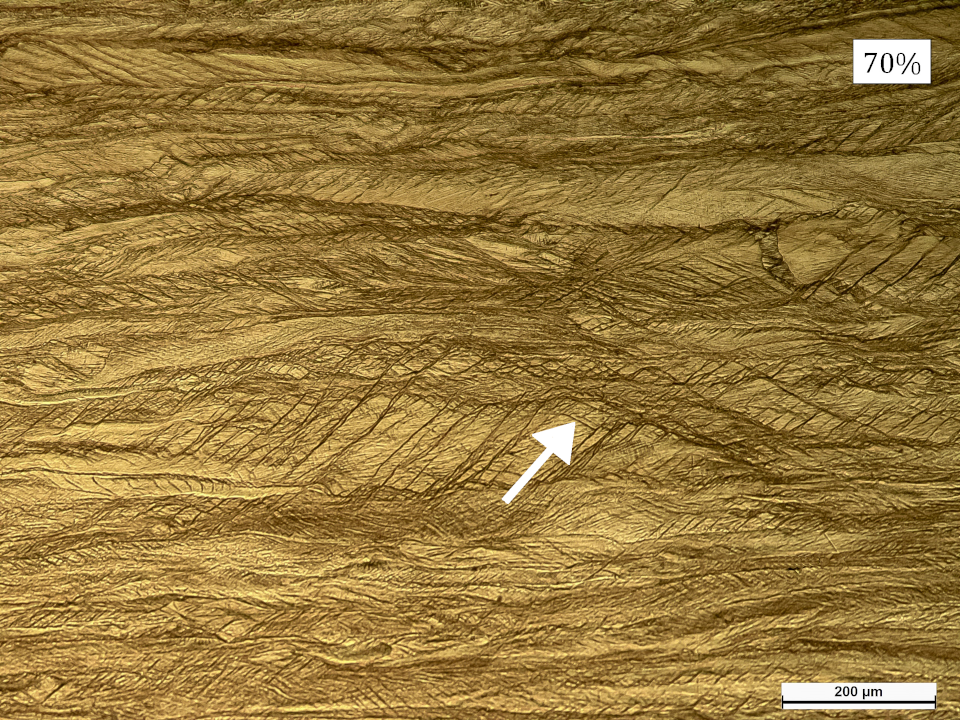}
\label{FIG:3.i}
\caption{Microstructures of cold rolled CuZn30 brass at (a) 0\%, (b) 10\%, (c) 15\%, (d) 20\%, (e) 30\%, (f) 40\%, (g) 50\%, (h) 60\%, and (i) 70\%  deformation. The observation plane is the RD-ND plane, with the rolling direction along the horizontal axis of the micrographs. Dashed arrows indicate strain marks and solid arrows indicate shear bands.}
\label{fIG:3}
\end{figure}

 Figure 4a illustrates that with increasing cold deformation, Vickers hardness shows a smooth rise characterized by two distinct changes in slope. Initially, the hardness increases with a relatively steep slope. However, beyond 30\% deformation, the hardness curve undergoes a change in slope, becoming less steep, indicating that the material's hardness continues to increase but at a slower rate than in the previous region.
 
Figure 4b depicts the evolution of dislocation density, calculated using equation (3). Initially, there is a consistent and gradual increase in dislocation density as deformation rises. This early stage of deformation is marked by a linear or nearly linear growth in dislocations, which aligns with the rise in Vickers Hardness observed in Figure 4a at this stage. However, an intriguing shift occurs at 30\% deformation. Beyond this point, the dislocation density experiences a much more significant increase compared to the initial stage. This pronounced rise in dislocation density suggests that the material might undergo a shift in its deformation mechanism or enter a new regime of plasticity.

\begin{figure}[ht!]
 \includegraphics[width=.49\textwidth]{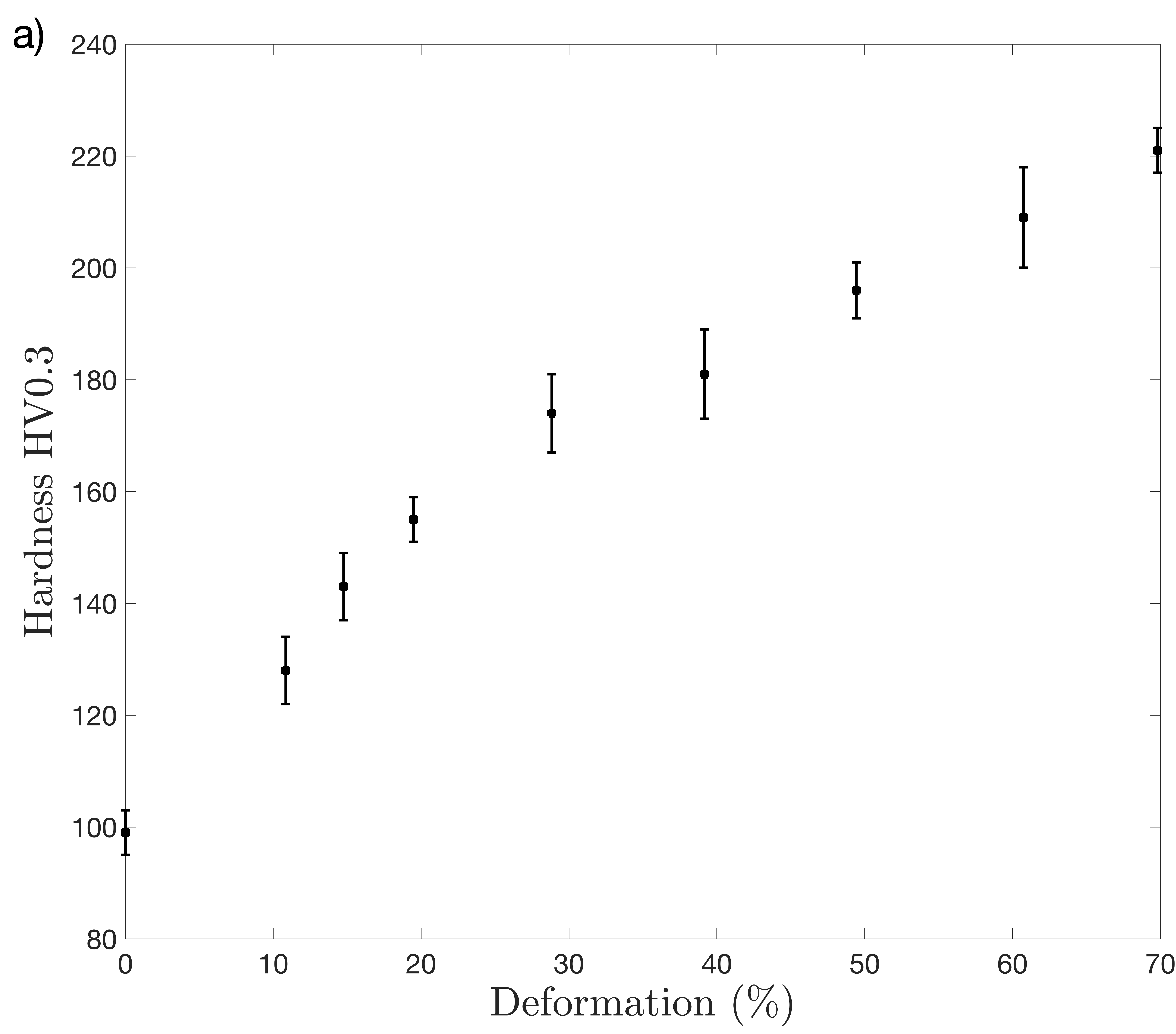}
\label{FIG:4.a}
 \includegraphics[width=.49\textwidth]{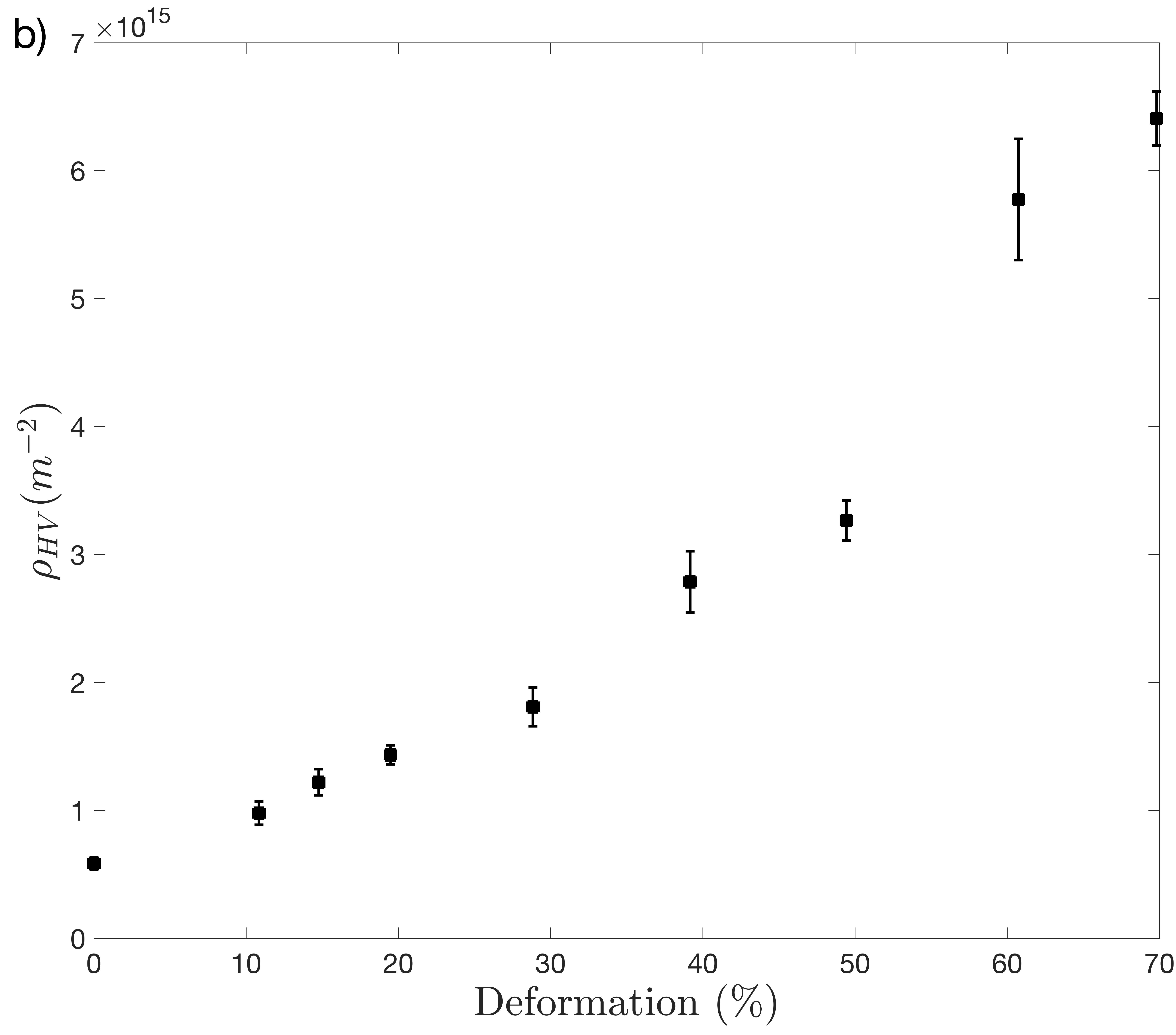}
\label{FIG:4.b}
\caption{a) Vickers Hardness and b) Evolution of dislocation density calculated by indentation method (equation (3)) vs. deformation.}
\label{FIG:4}
\end{figure}

\subsection{XRD results}\label{3.2}
Figure 5a show the normalized XRD patterns for cold rolled Cu30Zn brass, displaying broadening, displacement, and asymmetry of diffraction peaks due to deformation. Displacement and broadening are shown in more detail in Figure 5b, where the peak (111) is shown for the softened and 40\% deformation conditions. The above behavior is also observed for the peak (200) (Figure 5c), but asymmetry is mainly noticeable. It should be emphasized that displacement and asymmetry are a consequence of stacking faults and deformation twins; the competition between these sources of strain generated the variation of these peak aberrations \cite{ungar2004microstructural}.

Cell parameter, microstrain ($\langle\varepsilon^{2}\rangle^{1/2}$) and crystallite size (D) are values provided by the MAUD program after the refinement process. Cell parameter was 3.682 ± 0.001 \r{A}; microstrain and crystallite size are shown in Table 3, as well as the fitting parameter goodness of fit (GoFf) and as the weighted profile residual (Rwp), which indicate the fit of refinement processes. Microstrain constantly increases with a greater extent after 40\% deformation; the crystallite size, on the other hand, decreases sharply from the initial annealing condition to 10\% deformation and then decreases continuously to a lesser extent, which is consistent with the broadening of the x-ray diffraction peaks (Figure 5). 

Figure 6 shows Twinning fault probability (TFP) on the left axis, obtained directly from MAUD by considering Warren’s Model \cite{warren2012x}. TFP indicates the likelihood that a specific crystal will undergo twinning, and it implies that each successive layer (111) in the FCC sequence ABCABCABC should differ from the two preceding layers. Regions where the layers do not differ from the two preceding ones are identified as twin faults \cite{warren2012x}. Notably, TFP becomes significant after surpassing 20\% deformation, reaching its peak at 40\%. Afterward, it gradually declines and rises once more at the highest strain level considered. On the right axis, the dislocation density is plotted, calculated according to equation (4). It exhibits a continuous increase up to 40-50\% deformation, followed by a steeper rise, signifying an increase in the dislocation density rate. 

\begin{figure}[ht!]
	\centering
		\includegraphics[width=.9\textwidth]{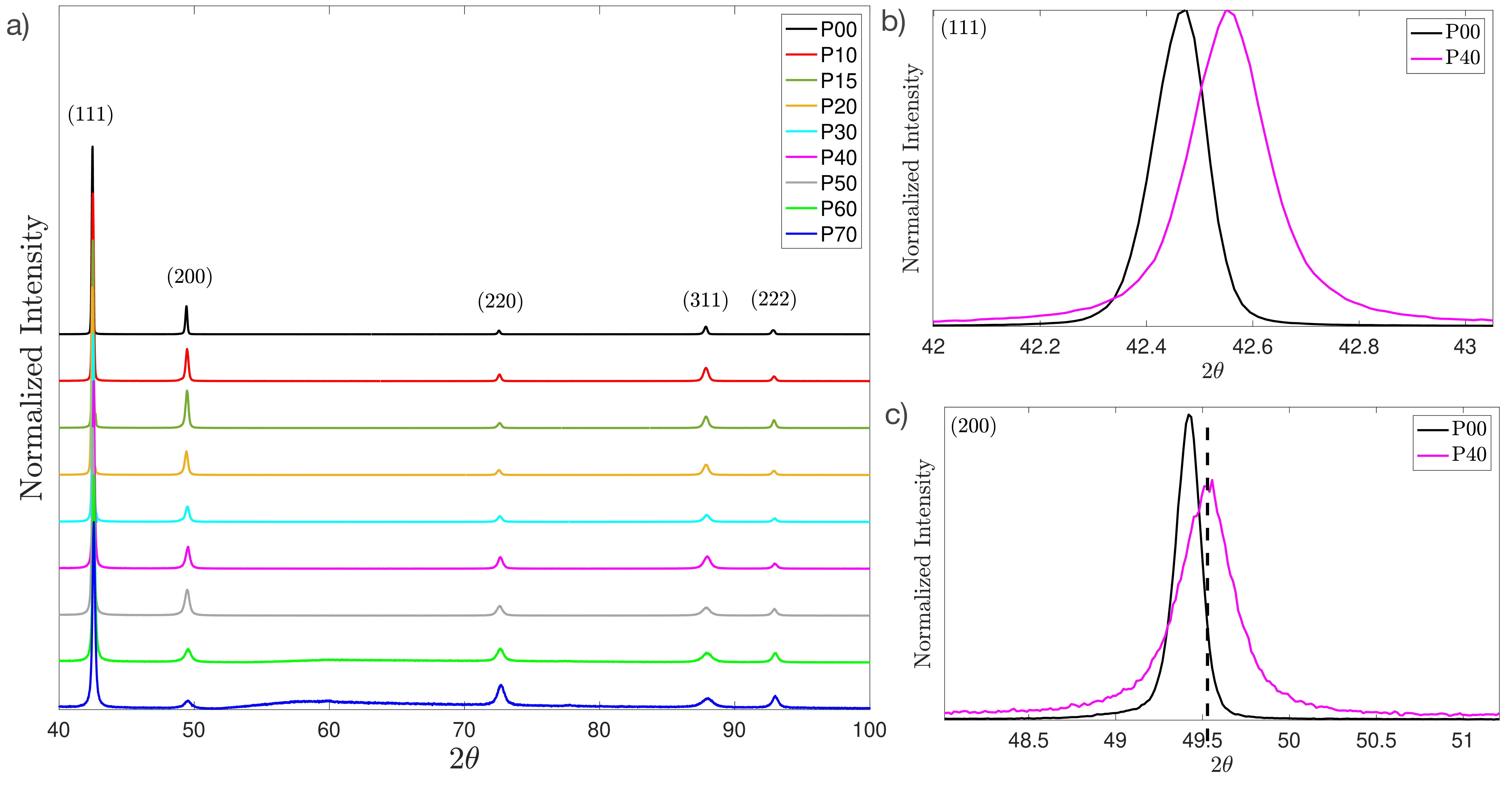}
	\caption{a) XRD pattern of cold rolled Cu30Zn brass at different deformations. Five peaks are observed corresponding to different lattice planes: (111) ($2\theta$ = 42.46), (200) ($2\theta$ = 49.42), (220) ($2\theta$ = 72.54), (311) ($2\theta$= 87.83) and (222) ($2\theta$ = 92.83). The curves go from top to bottom, increasing with the deformation. b) Peak (111) for the softened (P00) and 40\% deformation conditions (P40), showing displacement and broadening. c) Peak (200) for the softened (P00) and 40\% deformation conditions (P40) showing displacement and broadening and asymmetry.}
	\label{FIG:5}
\end{figure}

\begin{table}[width=.9\linewidth,cols=4,pos=h]
\begin{minipage}{\textwidth}
\renewcommand{\thefootnote}{\thempfootnote}
\caption{Microstructural characterization of the Cu30Zn rolled samples using the Rietveld Method.} 
\label{tbl3}
\begin{tabular*}
{\tblwidth}{@{} LLLLLLLLLL@{} }
\toprule
  & P00 & P10 & P15 & P20 & P30 & P40 & P50 & P60 & P70 \\
\midrule

microstrain x $10^{-03}$	& 0.838 & 1.185 & 1.121 & 1.284 & 1.250 & 1.324 & 1.432 & 1.830 & 2.168\\
microstrain (error) x $10^{-05}$ & 0.553 & 0.133 &0.739 & 0.521 & 2.540 & 3.830 & 2.109 & 1.059 & 2.267\\
Crystallite size (\r{A}) x $10^{03}$ & 11.488 & 2.105 & 1.774 & 1.529 & 0.950 & 0.694 & 0.607 & 0.459 & 0.042 \\
Crystallite size (\r{A}) (error) x $10^{01}$ & 77.040 & 3.919 & 1.968 & 1.572 & 1.059 & 0.964 & 0.717 & 3.212 & 0.602\\
GofF & 2.40 & 1.68 & 1.57 & 1.96 & 1.32 & 1.86 & 1.25 & 1.72 & 1.22\\
Rwp(\%) & 14.72 & 13.24 & 12.17 & 13.83 & 10.74 & 14.75 & 10.11 & 13.41 & 8.14\\

\bottomrule
 \end{tabular*}
 \end{minipage}
 \end{table}

\begin{figure}[ht!]
	\centering
		\includegraphics[width=.65\textwidth]{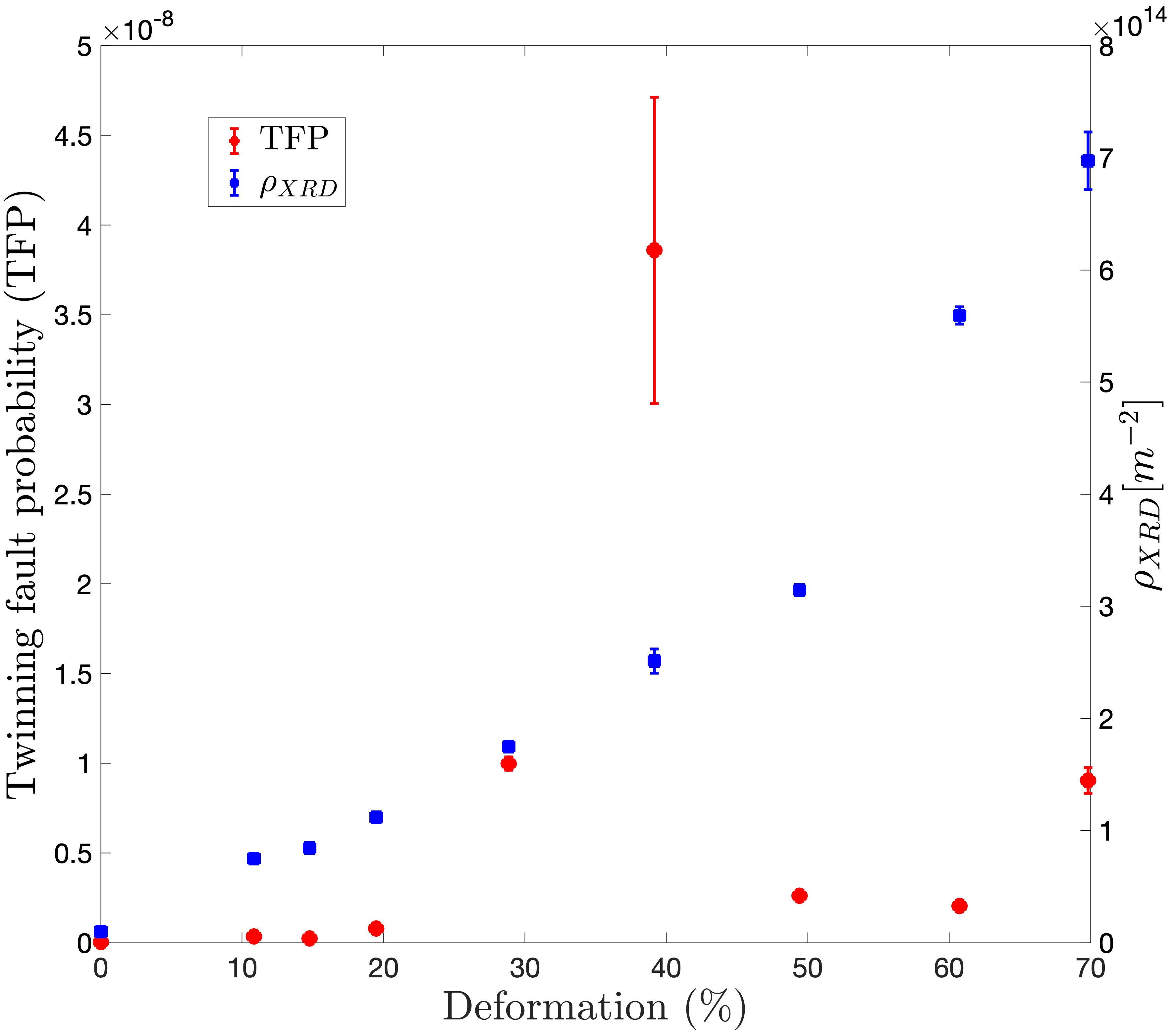}
	\caption{Twinning fault probability (TFP) (left axis) and dislocation density ($\rho_{XDR}$) (right axis) vs. deformation. Dislocation density was calculated according to equation (4) and TFP was obtained directly from MAUD by considering the Warren’s Model.}
	\label{FIG:6}
\end{figure}

\subsection{Ultrasound results}\label{3.3}

Linear ultrasound results are displayed in Figure 7a as the velocities of longitudinal wave ($V_{L,N}$), propagating along the thickness of the specimens; their percentage changes with deformation are shown in Table 3. Figure 7a shows that $V_{L,N}$ reaches a minimum value of 4540 m/s, that is, a 0.9\% decrease, at 20\% deformation and subsequently a maximum of 4684 m/s (2.24\% higher than the initial velocity) at 70\% deformation. The nonlinear acoustic parameter $\beta^{*}$ is shown in Figure 7b and its percentage change, in Table 3. $\beta^{*}$ reaches a first maximum value at 40\% deformation, about 737\% higher than in the annealed condition. $\beta^{*}$ subsequently decreases, remaining statistically constant  in the 50-60\% deformation range, and finally, increasing to 1280\% at the highest strain evaluated. It is important to notice that the large error bars of the nonlinear parameter are associated with the non-parallelism of the deformed specimens at the different measurement points and the high sensitivity of the parameter, generating some uncertainty between the different measurement points.

\begin{figure}[ht!]
 \includegraphics[width=.49\textwidth]{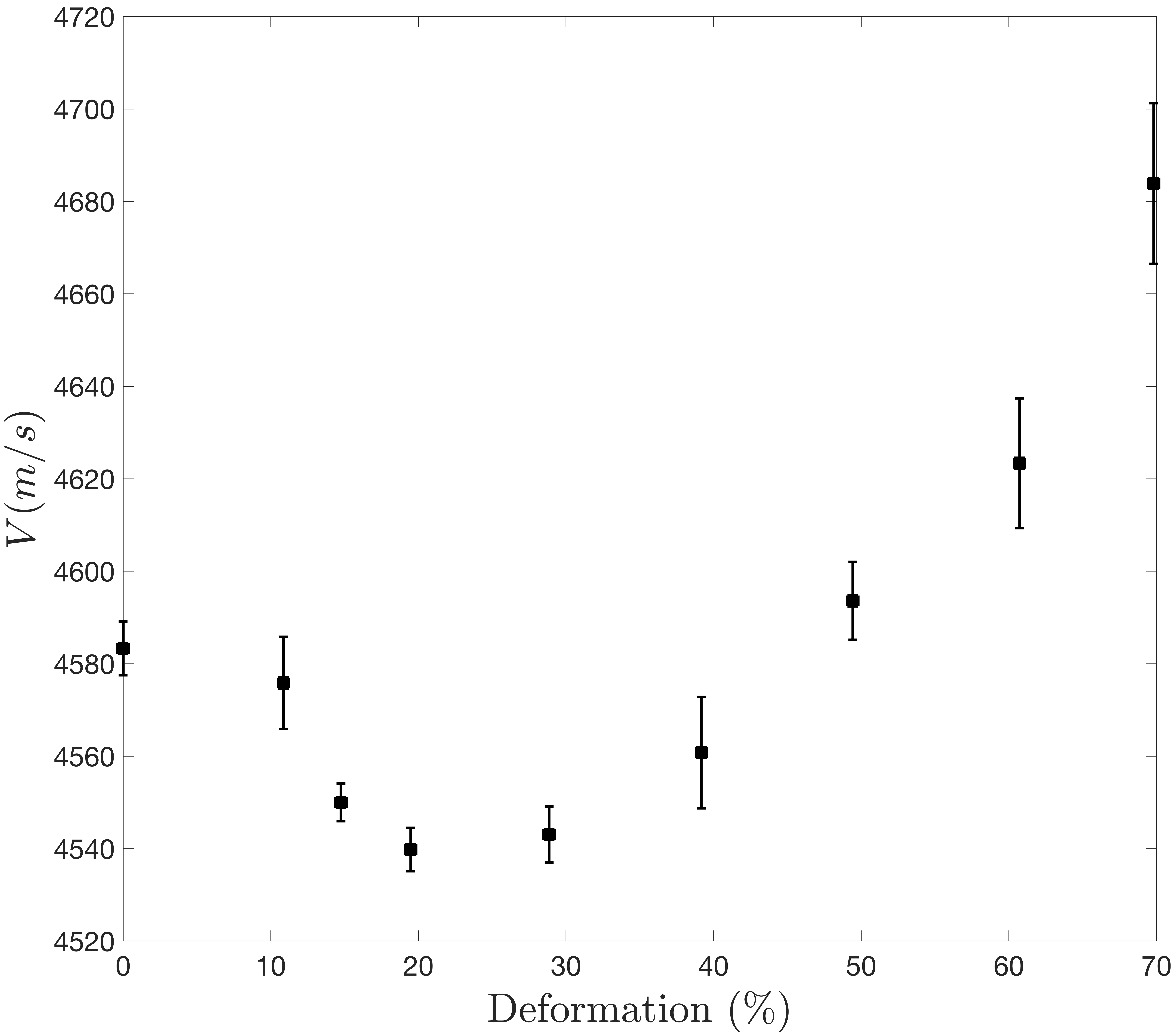}
\label{FIG:7.a}
 \includegraphics[width=.49\textwidth]{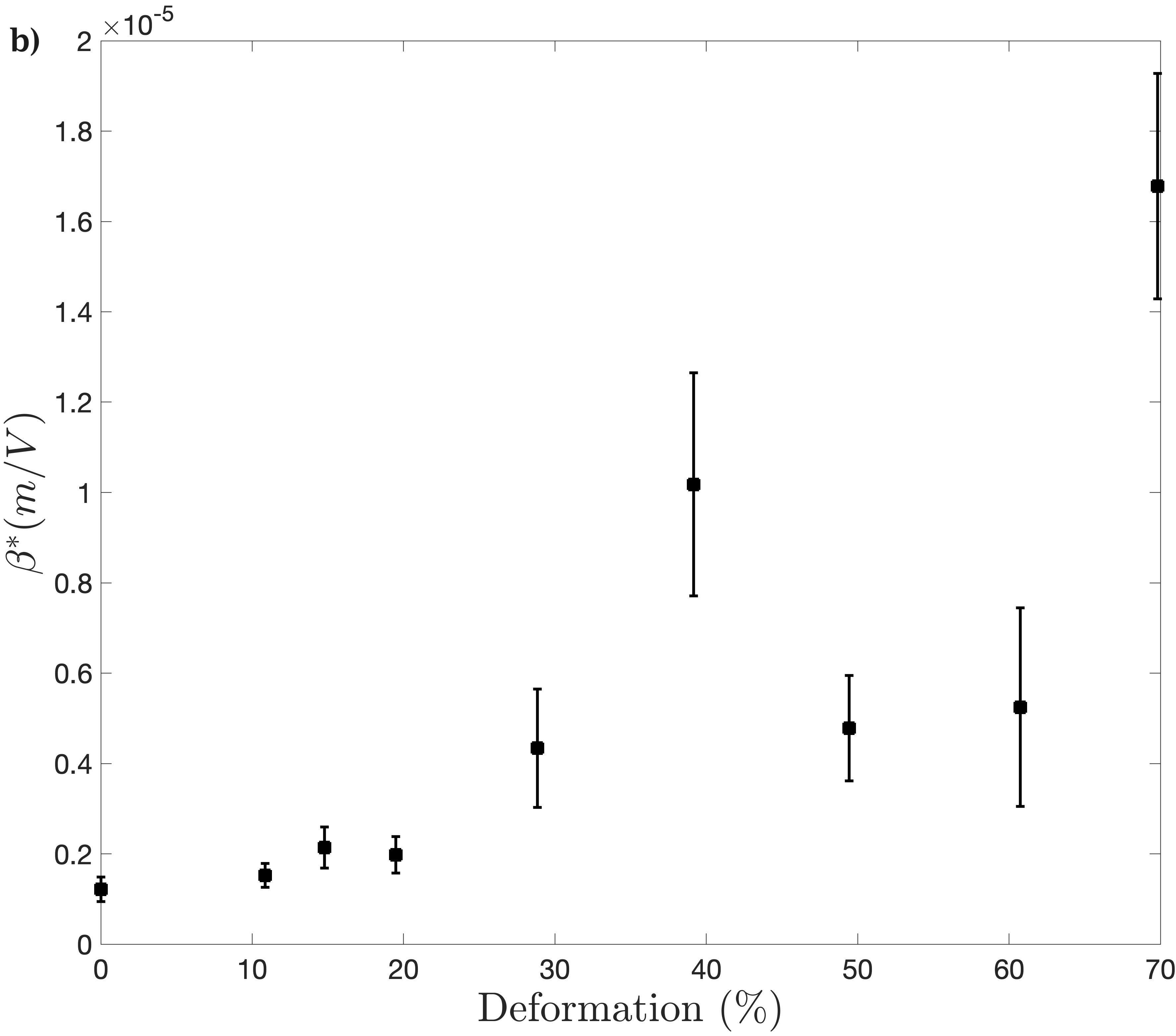}
\label{FIG:7.b}
\caption{a) Longitudinal wave velocity $V_{L,N}$ and b) Nonlinear parameter $\beta^{*}$ vs. deformation of cold-rolled Cu30Zn brass.}
\label{FIG:7}
\end{figure}

\begin{table}[width=.9\linewidth,cols=4,pos=h]
\caption{Percentage changes of linear and nonlinear acoustic parameters for cold-rolled Cu30Zn specimens at different deformations.} \label{tbl4}
\begin{tabular*}
{\tblwidth}{@{} LLLLLLLLLL@{} }
\toprule
   & P00 & P10 &P15 & P20 & P30 & P40 & P50 & P60 & P70\\
\midrule
$\Delta V_{L,N}$ (\%) & 0 & 0.12 & 0.68 & 0.90 & 0.83 & 0.44 & -0.27 & -0.92 & -2.24\\
$\Delta\beta^{*}$ (\%) & 0 & 25.27 & 76.18 & 62.74 & 257.04 & 737.38 & 293.30 & 331.57 & 1280.62\\
\bottomrule
\end{tabular*}
\end{table}

\section{Discussion} \label{4}
\subsection{ General Discussion}\label{4.1}

Authors \cite{fargette1976plastic, hirsch1985orientation, hutchinson1979development, madhavan2015role, duggan1978deformation} have reported an increment in strain marks as the deformation increases for cold deformed brasses. These marks may presumably be attributed to slip lines and bands or clusters of primary deformation twins that increase with the strain \cite{duggan1978deformation}. This result is corroborated with values obtained for TFP (Figure 6) which increases continuously from 10\% to 40\% deformation. After 40\% deformation, the microstructure shows the presence of shear bands (white arrow in figures 3g, 3h, and 3i). The formation of shear bands at 40\% deformation in cold rolled FCC metals has been reported elsewhere \cite{MORII1985379,anand2018correlation,HAASE2014327,BRACKE20091512,YAN2014408}, described as narrow zones angled at about 35° to the RD direction. The appearance of these shear bands coincides with the twinning saturation (TFP peak in Figure 6). Duggan et al. \cite{duggan1978deformation} explain that at this stage, dislocation slip is inhibited by the presence of an array of closely spaced twin boundaries, and the only slip systems which can continue to operate are those for which the glide plane is parallel to the twin plane; under these circumstances, the glide plane must rotate towards the rolling plane until the resolved shear stress is too small for a further slip of this type; as a consequence, the shear banding mechanism is the one that allows the deformation to continue. 

The observed formation of shear bands coincides with a drop in TFP. This observation could be explained by the detwinning process proposed by Hong et al. \cite{hong2010nucleation}. They describe the process of shear band formation as consisting of a nucleation state (bending, necking, and detwinning) and a thickening state. Nucleation of a shear band initiates in zones of localized deformation, bounded by a high density of dislocations resulting from strain accommodation. Necking of the twin lamellae starts at these boundaries by twin boundaries (TB) migration. The continuous migration with deformation entails breaking up and disappearing some TBs, a process called detwinning.

The statement by Hong et al. \cite{hong2010nucleation} further describes that by accumulating dislocations during the detwinning process, the subsequent shear strain can concentrate within shear bands, resulting in the formation of dislocation structures that evolve into a nano-sized sub-grain structure that is equiaxed or elongated along the shear direction. Other studies \cite{an2020deformation, wang2010detwinning} have shown that one of the significant consequences of the development of these shear bands is their role in transforming the twin–matrix (T-M) lamellar structure to nano-sized grains at large strains. This assertion will be of interest in the analysis of the following results.

In Figure 4a, the hardness results show two slight changes in slope, each corresponding to specific deformation intervals. Figure 4b, representing the dislocation density curve, also exhibits two intervals with changes in slope, which are inversely related to the hardness curve's slope changes. Remarkably, up to 40-50\% deformation, the dislocation density curve demonstrates a small slope, coinciding with the region where the hardness curve experiences its most significant hardening rate. This observation suggests that the material undergoes a relatively sluggish dislocation multiplication during this stage, indicating that dislocation interactions might not be the main mechanism contributing to the hardness. Instead, other mechanisms, such as twinning, could play a more crucial role in the observed hardening during this particular interval.

Once the material undergoes 40-50\% deformation, the slope of the dislocation density curve increases. Simultaneously, there is a minor reduction in the hardness curve. These findings imply that during this phase, the impact of twinning diminishes, and dislocation multiplication becomes the primary mechanism driving the hardening process.

The analysis of results obtained for dislocation density using XRD and Vickers hardness shows that hardness testing tends to overestimate the XRD-based dislocation density by approximately one order of magnitude. This is to be expected since the calculation of M, the Taylor factor, requires a detailed analysis of the fraction of texture components for each deformation to have a more precise value of M. The geometric factor $\alpha$ was estimated as 0.5, which has been described by the authors as overestimating the density of dislocations for low deformations. Likewise, hardness tests can cause the development of additional dislocations during the indentation process. However, is to be noted that the trend of the XRD-based dislocation density curve is fairly well described by the hardness approach, which makes it of value as a first semi-quantitative approach.

As previously discussed, the hardness test enables a comprehensive assessment of crystal imperfections over a larger area ($mm^{2}$) than XRD, which offers a more precise and detailed analysis of specific contributions like crystallite size (D) and deformation ($\langle\varepsilon^{2}\rangle^{1/2}$)  to dislocation density. The authors \cite{met11101571} have pointed out that using the indentation method to calculate dislocation density is an approximation, and certain factors need to be carefully taken into account to enhance the accuracy of such estimations. Nevertheless, despite its limitations, it serves as a valuable approach to illustrate the evolution of dislocation density in relation to hardness.

To analyze the effect of the microstructure on acoustic parameters, it must be considered that according to the theory of Hirsekorn \cite{10.1121/1.388233,10.1121/1.389206}, the wave velocity is inversely related to the grain size. An additional and opposite effect will come from the presence of dislocations \cite{PhysRevB.72.174110,PhysRevB.72.174111}: a higher dislocation density leads to smaller wave velocities.

The change in velocity of the longitudinal wave $V_{L,N}$ (Figure 7a), indicates that cold rolling reductionin alpha brass can be broadly divided into two stages. In the first stage, which is characterized by the action of slip and twinning, known as crystallographic deformation mechanisms \cite{el1999influence, el2000deformation, fargette1976plastic, hirsch1985orientation, hutchinson1979development, madhavan2015role, duggan1978deformation}, the effect of increasing dislocation density seems to prevail over twinning, resulting in a decrease in wave velocity up to 20\% deformation \cite{espinoza2018linear, salinas2022situ, PhysRevB.72.174110, PhysRevB.72.174111, salinas2017situ}. In the second stage, the increase in wave velocity could be attributed to twinning and subsequent formation of shear bands since both have a grain refining effect on microstructure.

\subsection{ Quantitative assessment of the effect of changes in dislocation density and in grain size upon the longitudinal waves velocities.}\label{4.2}.

According to the theory developed by Maurel et. al \cite{PhysRevB.72.174110,PhysRevB.72.174111} and verified with measurements on aluminum \cite{salinas2017situ}, steel and copper \cite{espinoza2018linear,salinas2022situ}, a change in dislocation density $\Delta\rho$ generates a change $\Delta V_{L,N}^{\rho}$  in longitudinal wave velocity $V_{L,N}$ given by

\begin{equation}
\centering
\frac{{\Delta V_{L,N}}^{\rho}}{V_{L,N_{0}}} = -\frac{32 L^{2}{\Delta\rho}}{15\pi^{4}\gamma^{2}}
\end{equation}

where $\gamma=\frac{V_{L}}{V_{S}}$ ($\gamma=2$ in our case) is the ratio between the longitudinal and shear waves velocities, and L is the mean distance between pinning points of the dislocations, modeled as vibrating strings of length L [27,28]. An immediate consequence is that an increase in dislocation density generates a decrease of the wave velocity. 

Similarly, the theory of Hirsekorn [45,46], indicates that a change $\Delta a^{2}$ in the mean grain size a of a polycrystal of cubic symmetry leads to a change $\Delta V_{L,N}^{g}$in the wave velocity $V_{L,N}$ given by

\begin{equation}
\centering
\frac{{\Delta V_{L,N}}^{g}}{V_{L,N_{0}}} = -p_{L} \kappa_{L}^{2} \Delta a^{2}
\end{equation}

where  $\kappa_{L}$ is the longitudinal wave number and $p_{L}> 0 $  is a dimensionless parameter to be determined by the experimental data. An immediate consequence can be drawn as well, that a decrease in grain size leads to an increase in wave velocity.

With 5 MHz, the longitudinal wavelength is about 1 mm. From Figure 3a the initial average grain size has been measured to be 45 microns, which gives $\kappa_{L}^{2}a_{0}^{2}\approx 8 \times 10^{-2}$. Figure 8 illustrates the wave velocities separately, determined from ultrasound results, $V_{L,N_{(US)}}$, and taking into account solely the influence of dislocations, $V_{L,N_{(\rho)}}$, and the variations in grain size, $V_{L,N_{(g)}}$. 

Based on the formulae (7) and (8), the following scenario emerges: Up to 20-30\% deformation the increase in dislocation density calculated by XDR, a value of L= 140 nm, and no significant change in the mean grain size make the $V_{L,N_{(\rho)}}$ fit well to $V_{L,N_{(US)}}$. Beyond that point there is a competition between the proliferation of dislocations, that tends to decrease the wave velocity, and a decrease in mean grain size, that tends to increase it. The resulting overall increase in $V_{L,N_{(US)}}$ from 20\% to 70\% deformation indicated in Figure 8 can be fit well with $p_{L}=2$ in $V_{L,N_{(g)}}$, assuming that dislocations continue to proliferate monotonously, as indicated by the hardness and XRD data, and that grains decrease in size by a factor of 3, also monotonically. In the early stages, say 20-50\%, this decrease will be dominated by twinning, while later on shear banding will come into play.

\begin{figure}[ht!]
\centering
\includegraphics[width=.65\textwidth]{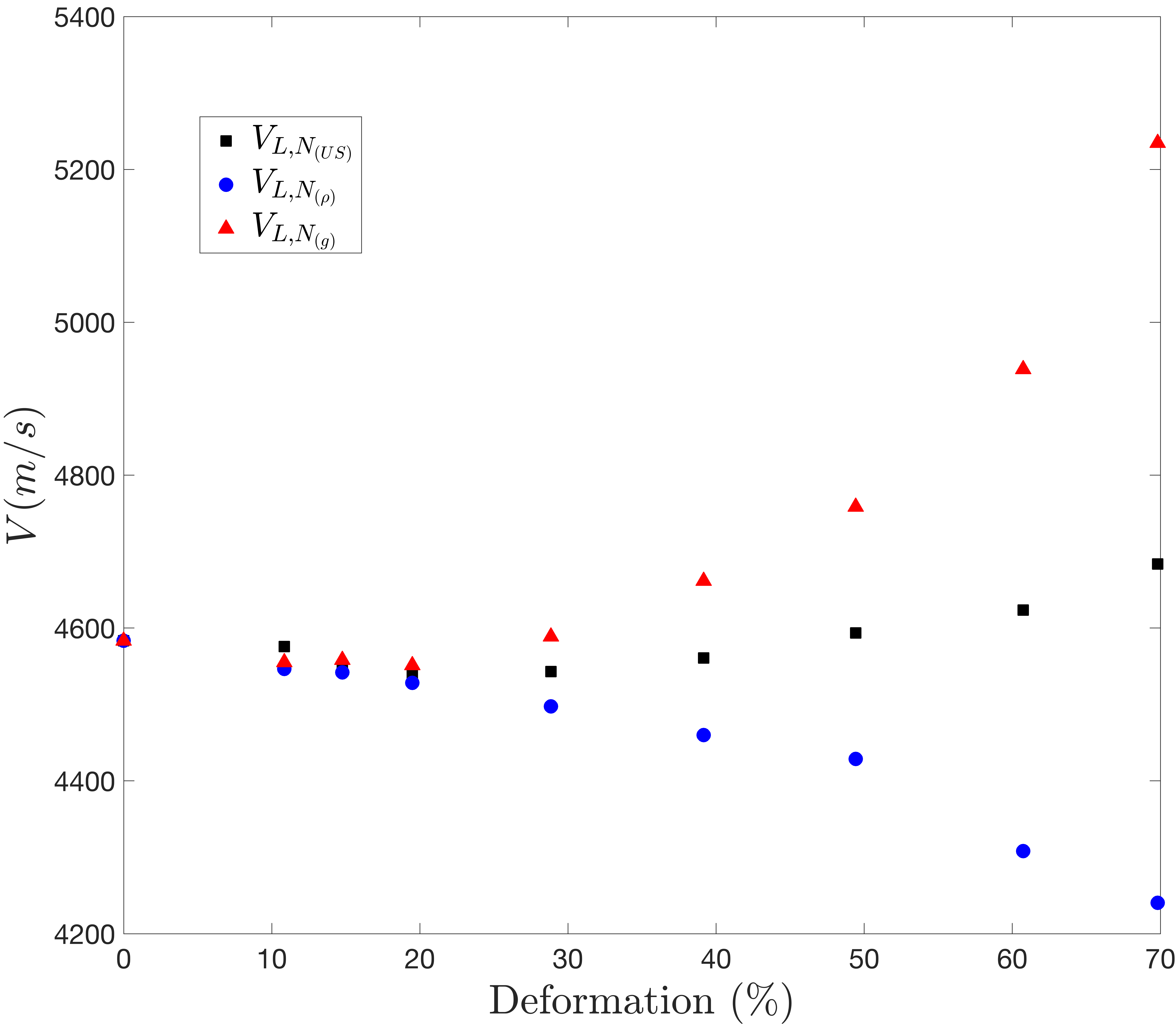}
\caption{Longitudinal wave velocity calculated by ultrasound results - $V_{L,N_{(US)}}$, by equation (7) - $V_{L,N_{(\rho)}}$ and by equation (8) - $V_{L,N_{(g)}}$ vs. deformation of cold-rolled Cu30Zn brass. The acoustic measurements (black squares) can be rationalized as a competition between a proliferation of dislocations (blue dots) and a decrease in average grain size (red triangles). }
\label{fig:8}
\end{figure}

\subsection{Overview}\label{4.3}

Based on the effect of microstructural features on longitudinal wave velocity, it is observed that $\beta^{*}$ increases at early stages of deformation due to two main factors: slip dislocation and deformation twinning. However, the data obtained from the XRD analysis suggests that twinning has a significant impact on the nonlinear parameter between 20-40\% deformation. During this strain interval, it is noted that there is a steady increase in $\rho_{XRD}$, while the substantial increase and saturation of twins possibly contributes to the considerable increase in $\beta^{*}$. 

The effect of decreasing grain size on acoustic parameters has been studied under the assumption that grain boundaries are discontinuities. Hence, a source of distortion of the ultrasonic wave due to the interception of the wave propagation path. Mini et al. \cite{mini2015experimental} investigated the effect of grain size on the nonlinear parameter, reporting that an increase in nonlinearity can be attributed to the increase in grain-boundary area and twin-boundary area. Therefore, twinning, which diminishes the effective grain size \cite{meyers2008mechanical}, leads to an increase in the nonlinearity parameter.

From a metallographic perspective, the continuous increase in longitudinal wave velocity and decrease in the nonlinear parameter after 40\% deformation could be attributed to shear band formation. Interestingly, while shear bands appear to have a similar effect on longitudinal wave velocity as deformation twinning, they exert a different impact on the nonlinear parameter. The formation of shear bands is considered a non-crystallographic mechanism as it involves breaking the microstructure without following any crystallographic relation. This implies a shift in the material's predominant deformation mechanism.

The microstructure inside shear bands is reported to consist of nano-sized sub-grain structures \cite{hong2010nucleation, an2020deformation,wang2010detwinning}, but the overall effect on acoustic parameters is a result of a combination of several factors, including:
\begin{itemize}
    \item A substantial increase in the dislocation density.
    \item The destruction of the twinning structure, leading to the emergence of new sites for twinning nucleation with further deformation \cite{hong2010nucleation, an2020deformation,wang2010detwinning}.
    \item Proliferation of shear bands.
    \item A significant component of Brass and Goss texture \cite{duggan1978deformation}.
\end{itemize}

Each of these factors exerts a distinct influence on wave propagation when considered individually. However, their collective interplay gives rise to a complex microstructure, with the effects balancing each other. The interplay of these various factors contributes to the observed changes in longitudinal wave velocity and the nonlinear parameter.

An in-depth comprehension of the microstructural alterations within the material and their influence on acoustic parameters is imperative for acquiring valuable insights into its mechanical characteristics. The interplay among shear bands, dislocation nucleation, twinning structures, and texture underscores the intricate nature of the material's response to deformation.

\section{Conclusions} \label{5}

The behavior of ultrasonic measurements in Cu30Zn brass has been correlated with the microstructure for deformation percentages up to 70\%. Microstructural analysis revealed an increase in strain marks from 10\% deformation and the formation of shear bands from 40\% deformation. Diffraction analysis further suggested the presence of deformation twins, reaching saturation at 40\% deformation. Simultaneously, calculated dislocation density exhibited a nearly linear increase during the early stages of deformation, followed by a significant rise after 40\% deformation, which coincided with the formation of shear bands.

While XRD provides a more comprehensive and precise analysis of individual contributions to dislocation density, hardness testing serves as a good approximation to demonstrate the evolution of dislocation density with hardness. However, to improve the accuracy of dislocation density estimations, various factors such as the Taylor factor and geometric factor alpha must be meticulously considered.

The impact of increased dislocations results in a 0.9\% decrease in wave velocity at 20\% strain. However, the combined effects of twinning and shear band formation lead to a significant increase in velocity, exceeding 2\% at 70\% deformation. Moreover, the simultaneous increase in dislocation and deformation twinning results in a remarkable rise of more than 700\% in the nonlinear parameter $\beta^{*}$ at 40\% deformation. Considering the constant increase in dislocation density within this deformation range, the substantial increase and saturation of twins likely play a crucial role in the considerable  rise of $\beta^{*}$.

Phenomena occurring after 40\% deformation, such as a significant increase in dislocation density rate, the reduction of the twinning structure, proliferation of shear bands, and the presence of Brass and Goss texture, collectively interact to generate a complex microstructure. Within this intricate microstructure, the effects of these various factors balance each other, leading to significant changes in longitudinal wave velocity and the nonlinear parameter. Disentangling the effect of these various factors on the behavior of the longitudinal wave velocity and the nonlinear parameter at high deformation remains a challenging task.

\section*{Acknowledgement}

This work was supported by Fondecyt Regular Grants \#11190900 (VS) and \#1230938 (FL).

%\bibliographystyle{unsrt}
%\bibliography{V_Beta_CuZn30}

\begin{thebibliography}{10}

\bibitem{asm1990properties}
ASM International.~Handbook Committee and American~Society for Metals.
\newblock {\em Properties and Selection: Nonferrous Alloys and Special-purpose
  Materials}, volume~2 of {\em ASM Handbook}.
\newblock ASM International, 1990.

\bibitem{Gallagher1970TheIO}
P.~Gallagher.
\newblock The influence of alloying, temperature, and related effects on the
  stacking fault energy.
\newblock {\em Metallurgical Transactions}, 1:2429--2461, 1970.

\bibitem{meyers2008mechanical}
M.A. Meyers and K.K. Chawla.
\newblock {\em Mechanical Behavior of Materials}.
\newblock Cambridge University Press, 2008.

\bibitem{el2000deformation}
E.~El-Danaf, S.~Kalidindi, R.~Doherty, and C.~Necker.
\newblock Deformation texture transition in brass: critical role of micro-scale
  shear bands.
\newblock {\em Acta Materialia}, 48(10):2665--2673, 2000.

\bibitem{el1999influence}
E.~El-Danaf, S.~Kalidindi, and R.~Doherty.
\newblock Influence of grain size and stacking-fault energy on deformation
  twinning in fcc metals.
\newblock {\em Metallurgical and Materials Transactions A}, 30(5):1223--1233,
  1999.

\bibitem{fargette1976plastic}
B.~Fargette and D.~Whitwham.
\newblock Plastic deformation of the brass \text{C}u30\text{Z}n by heavy
  rolling reductions.
\newblock {\em Mem. Sci. Rev. Metall.}, 73(3):197--206, 1976.

\bibitem{hirsch1985orientation}
J.~Hirsch, MY. Huh, and K.~L{\"u}cke.
\newblock Orientation dependence of the deformed microstructure in 70/30 brass.
\newblock {\em Strength of Metals and Alloys (ICSMA 7)}, pages 257--262, 1985.

\bibitem{hutchinson1979development}
WB. Hutchinson, BJ. Duggan, and M.~Hatherly.
\newblock Development of deformation texture and microstructure in cold-rolled
  \text{C}u--30\text{Z}n.
\newblock {\em Metals Technology}, 6(1):398--403, 1979.

\bibitem{madhavan2015role}
R.~Madhavan, R.~Kalsar, RK. Ray, and S.~Suwas.
\newblock Role of stacking fault energy on texture evolution revisited.
\newblock {\em IOP Conference Series: Materials Science and Engineering},
  82:012031, 2015.

\bibitem{duggan1978deformation}
BJ. Duggan, M.~Hatherly, WB. Hutchinson, and PT. Wakefield.
\newblock Deformation structures and textures in cold-rolled 70: 30 brass.
\newblock {\em Metal Science}, 12(8):343--351, 1978.

\bibitem{Kocks2003PhysicsAP}
U.~F. Kocks and H.~Mecking.
\newblock Physics and phenomenology of strain hardening: the fcc case.
\newblock {\em Progress in Materials Science}, 48(3):171--273, 2003.

\bibitem{Raphanel}
J.L. Raphanel.
\newblock {\em Large Plastic Deformations: Fundamental Aspects and Applications
  to Metal Forming}.
\newblock Routledge, London, 1st edition, 1993.

\bibitem{gupta}
V.~K. Gupta, N.~Tewary, M.~Yadav, and S.~K. Ghosh.
\newblock Effect of intercritical rolling on the microstructure, texture and
  mechanical properties of dual phase \text{TWIP} steel.
\newblock {\em Metallography, Microstructure, and Analysis}, 11(4):602--616,
  2022.

\bibitem{KUMARAN2023118814}
S.~N. Kumaran, S.~Sahoo, C.~Haase, L.~A. Barrales‐Mora, and L.~Tóth.
\newblock Nano-structuring of a high entropy alloy by severe plastic
  deformation: Experiments and crystal plasticity simulations.
\newblock {\em Acta Materialia}, 250:118814, 2023.

\bibitem{CHEN2022142136}
Y.~Chen, Y.~Liu, A.~Li, Z.~Ma, H.~Zhang, D.~Jiang, Y.~Ren, and L.~Cui.
\newblock Large-strain lüders-type deformation of b19' martensite in
  \text{N}i47\text{T}i49\text{N}b2\text{F}e2 alloy.
\newblock {\em Materials Science and Engineering: A}, 829:142136, 2022.

\bibitem{CHEN2022143224}
M.~Chen, J.~He, M.~Wang, J.~Li, S.~Xing, K.~Gui, G.~Wang, and Q.~Liu.
\newblock Effects of deep cold rolling on the evolution of microstructure,
  microtexture, and mechanical properties of 2507 duplex stainless steel.
\newblock {\em Materials Science and Engineering: A}, 845:143224, 2022.

\bibitem{DAN2018293}
C.~Dan, Z.~Chen, M.~Mathon, G.~Ji, L.~Li, Y.~Wu, F.~Brisset, L.~Guo, H.~Wang,
  and V.~Ji.
\newblock Cold rolling texture evolution of tib2 particle reinforced al-based
  composites by neutron diffraction and \text{EBSD} analysis.
\newblock {\em Materials Characterization}, 136:293--301, 2018.

\bibitem{burhan2019guideline}
I.~Burhan, G.~Mutaiyah, D.~Hashim, T.~Loganathan, and M.~Sultan.
\newblock A guideline of ultrasonic inspection on butt welded plates.
\newblock {\em IOP Conference Series: Materials Science and Engineering},
  554:012002, 2019.

\bibitem{shao2005review}
J.~Shao and Y.~Yan.
\newblock Review of techniques for on-line monitoring and inspection of laser
  welding.
\newblock {\em Journal of Physics: Conference Series}, 15:101--107, 2005.

\bibitem{merazi2010automatic}
T.~Merazi, B.~Boudraa, R.~Drai, and M.~Boudraa.
\newblock Automatic crack detection and characterization during ultrasonic
  inspection.
\newblock {\em Journal of Nondestructive Evaluation}, 29(3):169--174, 2010.

\bibitem{d2008automatic}
T.~D’orazio, M.~Leo, A.~Distante, C.~Guaragnella, V.~Pianese, and
  G.~Cavaccini.
\newblock Automatic ultrasonic inspection for internal defect detection in
  composite materials.
\newblock {\em NDT \& e International}, 41(2):145--154, 2008.

\bibitem{jhang2009nonlinear}
K.~Jhang.
\newblock Nonlinear ultrasonic techniques for nondestructive assessment of
  micro damage in material: a review.
\newblock {\em International journal of precision engineering and
  manufacturing}, 10(1):123--135, 2009.

\bibitem{kniazevnumerical}
V.~Kniazev.
\newblock Numerical investigation of acoustic nonlinearity for ultrasonic
  spectroscopy of interface defects in composites.
\newblock {\em Journal of Nondestructive Testing}, 16(5):1--9, 2011.

\bibitem{Buck1990}
Donald~O. Buck, O.Thompson and Dale~E. Chimenti.
\newblock {\em Nonlinear Acoustic Properties of Structural Materials. A
  Review}.
\newblock Springer, Boston, 1990.

\bibitem{cantrell2004fundamentals}
J.~Cantrell.
\newblock Fundamentals and applications of nonlinear ultrasonic nondestructive
  evaluation.
\newblock {\em Ultrasonic Nondestructive Evaluation: Engineering and biological
  material characterization}, 1:363--434, 2004.

\bibitem{carvajal2021ultrasonic}
L.~Carvajal, M.~Sosa, A.~Artigas, N.~Luco, and A.~Monsalve.
\newblock Ultrasonic assessment of the influence of cold rolling and
  recrystallization annealing on the elastic constants in a \text{TWIP} steel.
\newblock {\em Materials}, 14(21):65--59, 2021.

\bibitem{granato1956theory}
A.V. Granato and K.~L{\"u}cke.
\newblock Theory of mechanical damping due to dislocations.
\newblock {\em Journal of applied physics}, 27(6):583--593, 1956.

\bibitem{granato1956application}
A.V. Granato and K.~L{\"u}cke.
\newblock Application of dislocation theory to internal friction phenomena at
  high frequencies.
\newblock {\em Journal of applied physics}, 27(7):789--805, 1956.

\bibitem{hikata1965dislocation}
A.~Hikata, B.~Chick, and C.~Elbaum.
\newblock Dislocation contribution to the second harmonic generation of
  ultrasonic waves.
\newblock {\em Journal of Applied Physics}, 36(1):229--236, 1965.

\bibitem{hikata1966generation}
A.~Hikata and C.~Elbaum.
\newblock Generation of ultrasonic second and third harmonics due to
  dislocations. i.
\newblock {\em Physical Review}, 144(2):469--477, 1966.

\bibitem{cantrell1997effect}
J.~Cantrell and W.~Yost.
\newblock Effect of precipitate coherency strains on acoustic harmonic
  generation.
\newblock {\em Journal of applied physics}, 81(7):2957--2962, 1997.

\bibitem{cantrell1998nonlinear}
J.~Cantrell and X-G. Zhang.
\newblock Nonlinear acoustic response from precipitate-matrix misfit in a
  dislocation network.
\newblock {\em Journal of Applied Physics}, 84(10):5469--5472, 1998.

\bibitem{cantrell2006quantitative}
J.~Cantrell.
\newblock Quantitative assessment of fatigue damage accumulation in wavy slip
  metals from acoustic harmonic generation.
\newblock {\em Philosophical Magazine}, 86(11):1539--1554, 2006.

\bibitem{nazarov1997nonlinear}
V.~Nazarov and A.~Sutin.
\newblock Nonlinear elastic constants of solids with cracks.
\newblock {\em The Journal of the Acoustical Society of America},
  102(6):3349--3354, 1997.

\bibitem{166766}
D.~Hurley, D.~Balzar, P.~Purtscher, and K.~Hollman.
\newblock Nonlinear ultrasonic parameter in quenched martensitic steels.
\newblock {\em Journal of Applied Physics}, 83(9):4584--4588, 1998.

\bibitem{BALASUBRAMANIAM2011275}
K.~Balasubramaniam, Valluri J., and R.~Prakash.
\newblock Creep damage characterization using a low amplitude nonlinear
  ultrasonic technique.
\newblock {\em Materials Characterization}, 62(3):275--286, 2011.

\bibitem{li2019characterization}
W.~Li, B.~Chen, X.~Qing, and Y.~Cho.
\newblock Characterization of microstructural evolution by ultrasonic nonlinear
  parameters adjusted by attenuation factor.
\newblock {\em Metals}, 9(3):271, 2019.

\bibitem{viswanath2011nondestructive}
A.~Viswanath, B.~Rao, S.~Mahadevan, P.~Parameswaran, T.~Jayakumar, and B.~Raj.
\newblock Nondestructive assessment of tensile properties of cold worked
  \text{AISI} type 304 stainless steel using nonlinear ultrasonic technique.
\newblock {\em Journal of materials processing technology}, 211(3):538--544,
  2011.

\bibitem{espinoza2018linear}
C.~Espinoza, D.~Feli{\'u}, C.~Aguilar, R.~Espinoza-Gonz{\'a}lez, F.~Lund,
  V.~Salinas, and N.~Mujica.
\newblock Linear versus nonlinear acoustic probing of plasticity in metals: A
  quantitative assessment.
\newblock {\em Materials}, 11(11):2217, 2018.

\bibitem{matlack2015review}
K.~Matlack, J.~Kim, L.~Jacobs, and J.~Qu.
\newblock Review of second harmonic generation measurement techniques for
  material state determination in metals.
\newblock {\em Journal of Nondestructive Evaluation}, 34(1):273, 2015.

\bibitem{mini2015experimental}
R.~Mini, K.~Balasubramaniam, and P.~Ravindran.
\newblock An experimental investigation on the influence of annealed
  microstructure on wave propagation.
\newblock {\em Experimental Mechanics}, 55(6):1023--1030, 2015.

\bibitem{salinas2022situ}
V.~Salinas, C.~Aguilar, R.~Espinoza-Gonz{\'a}lez, J.~Gonz{\'a}lez,
  J.~Henr{\'\i}quez, F.~Lund, and N.~Mujica.
\newblock In-situ monitoring of dislocation proliferation during plastic
  deformation of 304\text{L} steel using ultrasound.
\newblock {\em Materials Science and Engineering: A}, 849:143416, 2022.

\bibitem{PhysRevB.72.174110}
A.~Maurel, V.~Pagneux, F.~Barra, and F.~Lund.
\newblock Interaction between an elastic wave and a single pinned dislocation.
\newblock {\em Phys. Rev. B}, 72:174110, 2005.

\bibitem{PhysRevB.72.174111}
A.~Maurel, V.~Pagneux, F.~Barra, and F.~Lund.
\newblock Wave propagation through a random array of pinned dislocations:
  Velocity change and attenuation in a generalized granato and l\"ucke theory.
\newblock {\em Phys. Rev. B}, 72:174111, 2005.

\bibitem{salinas2017situ}
V.~Salinas, C.~Aguilar, R.~Espinoza-Gonz{\'a}lez, F.~Lund, and N.~Mujica.
\newblock In situ monitoring of dislocation proliferation during plastic
  deformation using ultrasound.
\newblock {\em International Journal of Plasticity}, 97:178--193, 2017.

\bibitem{10.1121/1.388233}
S.~Hirsekorn.
\newblock The scattering of ultrasonic waves by polycrystals.
\newblock {\em The Journal of the Acoustical Society of America},
  72(3):1021--1031, 1982.

\bibitem{10.1121/1.389206}
S.~Hirsekorn.
\newblock The scattering of ultrasonic waves by polycrystals. ii. shear waves.
\newblock {\em The Journal of the Acoustical Society of America},
  73(4):1160--1163, 1983.

\bibitem{met11101571}
J.~Sidor, P.~Chakravarty, J.~Bátorfi, P.~Nagy, Q.~Xie, and J.~Gubicza.
\newblock Assessment of dislocation density by various techniques in cold
  rolled 1050 aluminum alloy.
\newblock {\em Metals}, 11(10), 2021.

\bibitem{DAN2018297}
M.~L. Taheri, H.~Weiland, and A.~D. Rollett.
\newblock A method of measuring stored energy macroscopically using
  statistically stored dislocations in commercial purity aluminum.
\newblock {\em Metall Mater Trans A}, 37(1):19–25, 2006.

\bibitem{SALEH2018620}
A.~A. Saleh, P.~Mannan, C.~N. Tomé, and E.~V. Pereloma.
\newblock On the evolution and modelling of cube texture during dynamic
  recrystallisation of \text{N}i–30\text{F}e–\text{N}b–\text{C} model
  alloy.
\newblock {\em Journal of Alloys and Compounds}, 748:620--636, 2018.

\bibitem{anthonyrecrystallization}
F.~Humphreys and M.~Hatherly.
\newblock {\em Recrystallization and Related Annealing Phenomena}.
\newblock Elsevier Science, 3rd edition, 2017.

\bibitem{HONG2003133}
S.~Hong and D.N. Lee.
\newblock The evolution of the cube recrystallization texture in cold rolled
  copper sheets.
\newblock {\em Materials Science and Engineering: A}, 351(1):133--147, 2003.

\bibitem{dieter1976mechanical}
G.~Dieter and D.~Bacon.
\newblock {\em Mechanical metallurgy}, volume~3.
\newblock McGraw-hill New York, 1976.

\bibitem{lutterotti1990simultaneous}
L.~Lutterotti and P.~Scardi.
\newblock Simultaneous structure and size-strain refinement by the rietveld
  method.
\newblock {\em Journal of applied Crystallography}, 23(4):246--252, 1990.

\bibitem{lutterotti1999maud}
L.~Lutterotti, S.~Matthies, and Hans R.
\newblock Maud: a friendly java program for material analysis using
  diffraction.
\newblock {\em IUCr: Newsletter of the CPD}, 21(14-15), 1999.

\bibitem{de1982use}
T.~De~Keijser, J.~Langford, E.~Mittemeijer, and A.~Vogels.
\newblock Use of the voigt function in a single-line method for the analysis of
  \text{X}-ray diffraction line broadening.
\newblock {\em Journal of Applied Crystallography}, 15(3):308--314, 1982.

\bibitem{delhez1993crystal}
R.~Delhez, T.~De~Keijser, J.~Langford, D.~Lou{\"e}r, E.~Mittemeijer, and
  E.~Sonneveld.
\newblock Crystal imperfection broadening and peak shape in the rietveld
  method.
\newblock {\em The Rietveld Method}, pages 132--166, 1993.

\bibitem{popa1998hkl}
N.~Popa.
\newblock The (hkl) dependence of diffraction-line broadening caused by strain
  and size for all laue groups in rietveld refinement: Journal of applied
  crystallography, 1998.

\bibitem{warren2012x}
B.~Warren.
\newblock {\em X-Ray Diffraction}.
\newblock Dover Books on Physics. Dover Publications, 2012.

\bibitem{bhaskar2014mechanical}
P.~Bhaskar, A.~Dasgupta, V.~S. Sarma, U.~K. Mudali, and S.~Saroja.
\newblock Mechanical properties and corrosion behaviour of nanocrystalline
  \text{T}i5\text{T}$a_{1.8}$\text{N}b alloy produced by cryo-rolling.
\newblock {\em Materials Science and Engineering: A}, 616:71--77, 2014.

\bibitem{american2004astm}
American~Society for Testing and Pennsylvania) Materials~(Filadelfia.
\newblock {\em \text{ASTM} E112-96(2004)e2: Standard Test Methods for
  Determining Average Grain Size}.
\newblock ASTM, 2004.

\bibitem{ungar2004microstructural}
T.~Ungar.
\newblock Microstructural parameters from x-ray diffraction peak broadening.
\newblock {\em Scripta Materialia}, 51(8):777--781, 2004.

\bibitem{MORII1985379}
K.~Morii, H.~Mecking, and Y.~Nakayama.
\newblock Development of shear bands in f.c.c. single crystals.
\newblock {\em Acta Metallurgica}, 33(3):379--386, 1985.

\bibitem{anand2018correlation}
K.~Anand, B.~Mahato, C.~Haase, A.~Kumar, and S.~Chowdhury.
\newblock Correlation of defect density with texture evolution during cold
  rolling of a twinning-induced plasticity (\text{TWIP}) steel.
\newblock {\em Materials Science and Engineering: A}, 711:69--77, 2018.

\bibitem{HAASE2014327}
C.~Haase, L~A. Barrales‐Mora, F.~Roters, D.~A. Molodov, and G.~Gottstein.
\newblock Applying the texture analysis for optimizing thermomechanical
  treatment of high manganese twinning-induced plasticity steel.
\newblock {\em Acta Materialia}, 80:327--340, 2014.

\bibitem{BRACKE20091512}
L.~Bracke, K.~Verbeken, L.~Kestens, and J.~Penning.
\newblock Microstructure and texture evolution during cold rolling and
  annealing of a high \text{M}n twip steel.
\newblock {\em Acta Materialia}, 57(5):1512--1524, 2009.

\bibitem{YAN2014408}
Y.~Haile, Z.~Xiang, J.~Nan, Z.~Yiran, and H.~Tong.
\newblock Influence of shear banding on the formation of brass-type textures in
  polycrystalline fcc metals with low stacking fault energy.
\newblock {\em Journal of Materials Science \& Technology}, 30(4):408--416,
  2014.

\bibitem{REN2021111013}
P.~Ren, X.P. Chen, C.Y. Wang, Y.X. Zhou, W.Q. Cao, and Q.~Liu.
\newblock Evolution of microstructure, texture and mechanical properties of
  \text{F}e–30\text{M}n–11\text{A}l–1.2\text{C} low-density steel during
  cold rolling.
\newblock {\em Materials Characterization}, 174:111013, 2021.

\bibitem{hong2010nucleation}
C.~Hong, N.~Tao, X.~Huang, and K.~Lu.
\newblock Nucleation and thickening of shear bands in nano-scale twin/matrix
  lamellae of a \text{C}u-\text{A}l alloy processed by dynamic plastic
  deformation.
\newblock {\em Acta Materialia}, 58(8):3103--3116, 2010.

\bibitem{an2020deformation}
X.~An, S.~Ni, M.~Song, and X.~Liao.
\newblock Deformation twinning and detwinning in face-centered cubic metallic
  materials.
\newblock {\em Advanced Engineering Materials}, 22(1), 2020.

\bibitem{wang2010detwinning}
J.~Wang, N.~Li, O.~Anderoglu, X.~Zhang, A.~Misra, J.~Huang, and J.~Hirth.
\newblock Detwinning mechanisms for growth twins in face-centered cubic metals.
\newblock {\em Acta Materialia}, 58(6):2262--2270, 2010.

\bibitem{KONKOVA2015173}
T.~Konkova, S.~Mironov, A.~Korznikov, G.~Korznikova, M.M. Myshlyaev, and S.L.
  Semiatin.
\newblock An \text{EBSD} investigation of cryogenically-rolled
  \text{C}u–30\%\text{Z}n brass.
\newblock {\em Materials Characterization}, 101:173--179, 2015.

\end{thebibliography}

\end{document}